\documentclass[sigconf]{acmart}
\usepackage{pdfpages}
\usepackage[inkscapelatex=false]{svg}
\usepackage{pbox}
\usepackage{algorithm}
\usepackage{algpseudocode}

\usepackage{mathtools}

\usepackage{physics}
\usepackage{tikz}
\usepackage{subcaption}

\newcommand{\sample}{\overset{\$}{\leftarrow}}
\newcommand{\smallsubsec}[1]{\vspace{0.01in}\smallskip\noindent\textbf{#1: }}
\renewcommand{\vec}[1]{\mathbf{#1}}
\newcommand{\oper}[1]{\mathsf{{#1}}}

\newcommand{\key}[1]{\texttt{#1}}

\newcommand{\bsk}{\key{BSK}}
\newcommand{\ksk}{\key{KSK}}
\newcommand{\sk}{\key{sk}}
\newcommand{\Sk}{\key{Sk}}

\newcommand{\decomp}{\oper{Decomp}^{\beta, \ell}}
\newcommand{\dft}{\oper{DFT}}
\newcommand{\idft}{\oper{DFT}^{-1}}

\newcommand{\hatt}[1]{\hat{\hat{#1}}}

\newcommand{\ct}[1]{\text{#1}}

\newcommand{\rlwe}{\ct{RLWE}}
\newcommand{\lev}{\ct{Lev}}

\newcommand{\rgsw}{\ct{RGSW}}

\newcommand{\throu}{\mathcal T}
\newcommand{\HRule}{\noindent\rule{\linewidth}{0.4pt}}

\setcopyright{none} 

\author{Valentin Reyes Häusler, Gabriel Ott, Aruna Jayasena, Andreas Peter \newline \{valentin.reyes.haeusler, gabriel.ott, andreas.peter\}@uni-oldenburg.de, aruna@tennessee.edu}

\begin{document}
\title{Towards a Functionally Complete and Parameterizable\\ TFHE Processor}


\begin{abstract}
Fully homomorphic encryption allows the evaluation of arbitrary functions on encrypted data. It can be leveraged to secure outsourced and multiparty computation. TFHE is a fast torus-based fully homomorphic encryption scheme that allows both linear operations, as well as the evaluation of arbitrary non-linear functions. It currently provides the fastest bootstrapping operation performance of any other FHE scheme.
Despite its fast performance, TFHE suffers from a considerably higher computational overhead for the evaluation of homomorphic circuits. Computations in the encrypted domain are orders of magnitude slower than their unencrypted equivalents. This bottleneck hinders the widespread adoption of (T)FHE for the protection of sensitive data.

While state-of-the-art implementations focused on accelerating and outsourcing single operations, their scalability and practicality are constrained by high memory bandwidth costs. In order to overcome this, we propose an FPGA-based hardware accelerator for the evaluation of homomorphic circuits. Specifically, we design a functionally complete TFHE processor for FPGA hardware capable of processing instructions on the data completely on the FPGA. In order to achieve a higher throughput from our TFHE processor, we implement an improved programmable bootstrapping module which outperforms the current state-of-the-art by 240\% to 480\% more bootstrappings per second. Our efficient, compact, and scalable design lays the foundation for implementing complete FPGA-based TFHE processor architectures\footnote{Open access: \url{https://github.com/CeresB/tfhe-processor-artifacts}}.

\end{abstract}


\maketitle

\section{Introduction}\label{sec:intro}
Secure multiparty computation aims to enable several parties to jointly perform computations over sensitive data without any party divulging their personal secret inputs. The homomorphic encryption approach relies on the construction of encryption schemes that allow the evaluation of computations on data while in encrypted form. While most such schemes are constrained to only a handful of operations, the branch of fully homomorphic encryption (FHE) specifically aims to allow any arbitrary computation on arbitrary encrypted data. 


The first successful construction of an FHE scheme was accomplished by Gentry~\cite{gentry09} in 2009. This was followed by a large body of works leveraging and improving on Gentry's blueprint, leading to faster and more practical fully homomorphic encryption schemes. Despite these optimizations, FHE suffers from a considerable computational overhead when evaluating encrypted computations. Concretely, FHE schemes are several orders of magnitude slower than equivalent plaintext computations~\cite{entry_whitepaper}. This severely hinders the adoption of FHE as a general solution for secure multiparty computation. 

The torus-based encryption scheme TFHE~\cite{TFHE_original_paper} leverages the learning with errors (LWE) hardness assumption~\cite{Regev2005} to provide fully homomorphic encryption. TFHE relies on small parameter sets in order to provide fast homomorphic operations. Additionally, the bootstrapping operation allows the inherent encryption noise to be kept within secure and correct bounds. This is typically the most expensive operation in FHE schemes. TFHE accomplishes this by providing the fastest FHE bootstrapping algorithm, while also allowing the evaluation of arbitrary lookup tables for free during bootstrapping. This is considerable as it allows the evaluation of non-linear operations under encryption, leading to the term programmable bootstrapping (PBS).

Several works have explored hardware acceleration for FHE, particularly TFHE~\cite{ALT, Tian_Ye, Tianqi_Kong, FPT}. Most focus on accelerating the PBS operation on FPGAs, offloading only this computationally intensive step while keeping other homomorphic operations on conventional hardware. Although this design enables significant PBS acceleration, it introduces high data transfer overheads to and from the FPGA, making overall performance and scalability heavily dependent on external memory bandwidth.

\begin{figure}
    \centering
    \input{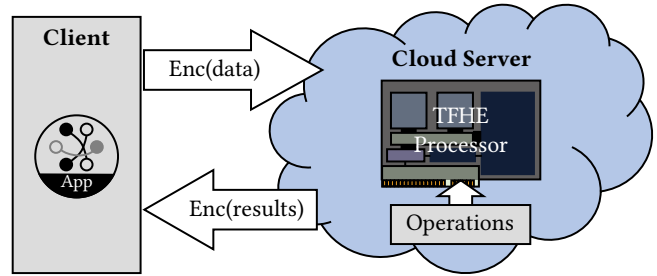}
    \vspace{-0.2in}
    \caption{Encrypted data alongside operations are passed to the processor in order to evaluate a homomorphic circuit.}
    \label{fig:basic_idea}
    \vspace{-0.3in}
\end{figure}

In this paper, we present a fully FPGA-based TFHE accelerator that performs the programmable bootstrapping (PBS) operation entirely on the FPGA, as illustrated in Figure~\ref{fig:basic_idea}. To eliminate memory bandwidth bottlenecks, we develop a complete FPGA-based TFHE processor centered around a compact and efficient PBS module. Our design achieves 240\% to 480\% more PBS operations per second compared to the current state of the art. This resource-efficient design enables the integration of an additional key-switch module, providing full TFHE functionality. Moreover, the proposed key-switch unit supports both addition and scalar multiplication, completing the set of operations required for TFHE.



\subsection{Related Work}

In this section, we review prior works that employ FPGA or dedicated hardware accelerators for implementing TFHE processors.

\begin{table*}
    \caption{Comparison of prior work. FPT and ALT accelerate the blind rotation but do not perform a complete PBS as only the blind rotation step is performed on the FPGA~\cite{FPT,ALT}. Abbreviations: BR (Blind rotation), KS (Keyswitch)}
    \label{tab:related_works}
    \centering
    \begin{tabular}{c|c|c|c|c|c|c|c|l}\hline
        \\[-1em] 
        Work & \pbox{1cm}{DFT Alg.} & \pbox{1.5cm}{NTT prime size in bits} & \pbox{0.5cm}{BR} & \pbox{0.5cm}{PBS} & \pbox{1.2cm}{Key unrolling}& \pbox{1.6cm}{Rotation method} & \pbox{1.4cm}{Bottleneck} & Other\\ \hline
        \\[-1em] 
        FPT~\cite{FPT} & FFT & - & \checkmark & - & no & memory & computational & \pbox{4.6cm}{Batched bootstrapping}\\ \hline
        \\[-1em] 
        ALT~\cite{ALT} & NTT & \pbox{1cm}{17 \& 16} & \checkmark & - & yes & multiplication & computational & \pbox{4.6cm}{Composite NTT}\\ \hline
        \\[-1em] 
        KL~\cite{Tianqi_Kong} & NTT & 52 & \checkmark & \checkmark & yes & memory & bandwidth & \pbox{4.6cm}{Invalid keyswitch implementation}\\ \hline
        \\[-1em] 
        YKP~\cite{Tian_Ye} & NTT & 33 & \checkmark & \checkmark & yes & multiplication & bandwidth & \pbox{4.7cm}{-} \\ \hline
        \\[-1em] 
        MXP~\cite{MXP} & FFT & - & \checkmark & \checkmark & N/A & N/A & N/A & \pbox{4.6cm}{Small-scale CPU-like architecture with special instructions to accelerate common TFHE operations}\\ \hline
        \\[-1em] 
        Ohba~\cite{Ohba2025_nvme} & NTT & - & \checkmark & \checkmark & N/A & N/A & N/A & \pbox{4.6cm}{CPU-like architecture with special instruction to accelerate NTT computation}\\ 
    \end{tabular}

\end{table*}

\smallsubsec{TFHE Acceleration with FPGA} 
There are several promising efforts that aim to accelerate TFHE by leveraging FPGAs, as summarized by Table~\ref{tab:related_works}. In this section, we provide an overview of these works alongside their major limitations.

Gener et al.~\cite{MXP} propose an accelerated general-purpose vector engine and demonstrate its capabilities by accelerating the polynomial multiplication used for TFHE bootstrapping. Their solution is not specific to TFHE and performs considerably slower than comparable CPU implementations. Ohba et al.~\cite{Ohba2025_nvme} similarly present a general-purpose NTT acceleration engine, which is then applied to TFHE. Similarly, Turbo-FHE~\cite{Turbo_FHE} leverages the Karatsuba algorithm to accelerate polynomial multiplication on FPGAs. The authors do not provide sufficient details for a comparison, nor do they provide comparisons themselves. 
FPT~\cite{FPT} accelerates polynomial multiplication by using fixed-point arithmetic for the Fast Fourier Transform (FFT). However, this approach limits numerical precision, increasing the probability of decryption failures. Similar to our design, FPT leverages on-chip High-Bandwidth Memory (HBM) to mitigate external memory bandwidth constraints. Furthermore, its batched operation scheduling provides additional time between memory accesses, effectively reducing overall bandwidth demand.

ALT~\cite{ALT}, Kong and Li (KL)~\cite{Tianqi_Kong}, as well as Ye et al. (YKP)~\cite{Tian_Ye}, accelerate the programmable bootstrapping operation by leveraging key-unrolling and performing several small NTTs in parallel. Additionally, ALT uses two small primes to perform composite NTTs, leveraging the Chinese remainder theorem. All three works rely on external off-chip memory to store the bootstrapping key, adding large memory bandwidth constraints. ALT solves this problem by prefetching parts of the key during computation. The key-switching operation implemented by Kong and Li requires the secret key to be present in the clear on the FPGA, compromising the security of the implementation.

\smallsubsec{FHE Processors beyond TFHE} 
Several promising efforts have explored dedicated FHE processor designs for schemes beyond TFHE. Specifically, BASALISC~\cite{BASALISC} targets the BGV scheme, while F1~\cite{F1} introduces a general-purpose FHE processor compatible with BGV, CKKS, and general polynomial operations~\cite{CoFHEE_extended}. These efforts present ASIC-based designs emphasizing efficient memory management, hardware reuse for compactness, and support for diverse FHE parameter sets. Each design also defines its own scheme-specific FHE instruction set. However, these designs remain limited in their applicability to TFHE, as their architectures and instruction sets are optimized for FHE schemes which function fundamentally different to TFHE regarding their parameters and capabilities.



\section{Preliminaries}
\label{sec:preliminaries}

In this section, we first introduce the notation used throughout the paper, followed by an overview of the TFHE scheme and FPGA-based hardware acceleration.

\subsection{Notation}
We denote with $\mathbb B = \{0,1\}$ the binary set, with $\mathbb Z_q$ the set of integers modulo $q$, and with $\mathbb T = \mathbb R / \mathbb Z$ the torus of real numbers modulo~$1$. For a set $S$, we denote with $S[X]$ the set of polynomials with coefficients in $S$, and with $S_N[X] = S[X]/(X^N + 1)$ the set of polynomials in $S[X]$ with degree $< N$. More concretely, we define $\mathfrak R_q^N = \mathbb Z_q[X] / (X^N + 1)$ the quotient ring of polynomials of degree $< N$ and coefficients in $\mathbb Z_q$. Note that $\mathbb Z_q \cong \frac{1}{q} \mathbb Z / \mathbb Z \subset \mathbb T$, i.e. $\mathbb Z_q$ is isomorphic to the discretized torus $\frac{1}{q} \mathbb Z / \mathbb Z$.

For a distribution $D$, we denote with $e \leftarrow D$ the sampling of an element $e$ according to the distribution $D$. Similarly, for a set $S$ we denote with $e \sample S$ the uniformly random sampling of $e$ from $S$. We denote with $\chi_q$ a zero-centered Gaussian distribution with standard deviation $\sigma$.

We use bold letters to denote vectors $\vec v$. The angled bracket notation $\langle \vec a, \vec b \rangle$ is used to denote the dot-product of two equal-length vectors. The element-wise multiplication of two vectors is denoted with $\vec a * \vec b$. For a vector $\vec v$ we denote with $|| \vec v||_\infty$ the infinity norm. We naturally extend this notation to polynomials by considering the representation as vectors of coefficients. The notation $\lfloor x \rceil$, $\lfloor x \rfloor$ and $\lceil x \rceil$ respectively denote the rounding to the nearest integer, rounding down, and rounding up of $x$. 
We denote with $\throu$ the throughput of an FPGA module. All logarithms are computed to base 2.

\subsection{Torus-based Fully homomorphic encryption}\label{sec:tfhe}
Fully homomorphic encryption (FHE) schemes enable the evaluation of arbitrary functions on encrypted data. This is typically accomplished by constructing schemes that allow both unlimited additions and multiplications under encryption, thus providing functional completeness. All major FHE schemes rely on noise for encryption. This noise grows with each homomorphic operation, reducing the probability of correct decryption. The introduction of the \emph{bootstrapping} operation to manage this noise led to the construction of the first FHE scheme by Gentry~\cite{gentry09}.

The vast majority of FHE schemes rely on the learning with errors (LWE) hardness assumption~\cite{Regev2005}, and its ring-variant RLWE~\cite{lyubashevsky_ideal_2012}, for encryption. The generalized LWE (GLWE) problems capture both variants as illustrated by Definition~\ref {def:GLWE}.

\HRule
\begin{definition} \label{def:GLWE}(GLWE~\cite{Regev2005,lyubashevsky_ideal_2012})
Let $k \geq 1$ be an integer and $N$ a power of two and $X^N + 1$ the corresponding cyclotomic polynomial. Let $\chi$ be a Gaussian distribution over $\mathbb T_N[X]$ with coefficient-wise variance $\sigma^2$, and let $s \in \mathbb B_N[X]$. Let $\text{GLWE}_{s, \chi}$ denote the distribution over $\mathbb T_N[X]^k \times \mathbb T_N[X]$ obtained by sampling a uniform vector $\vec a \sample \mathbb T_N[X]^k$ and yielding the tuple $(\vec a, b)$ with $b = \langle \vec a, \vec s \rangle + e$, where $e \leftarrow \chi$.
\begin{description}
    \item[Search problem:] Given arbitrary many independent samples from $\text{GLWE}_{s, \chi}$, find $s$.
    \item[Decision problem:] Given arbitrary many tuples $(\vec a, b) \in \mathbb T[X]^k \times \mathbb T[X]$, determine whether the tuples were sampled uniformly at random from $\mathbb T[X]^k \times \mathbb T[X]$ or sampled from $\text{GLWE}_{s, \chi}$.
\end{description}
\HRule
\end{definition}

Both problems in Definition~\ref {def:GLWE} are assumed to be hard even against quantum computing algorithms. By choosing $N = 1$ the above defines the LWE hardness assumption. For $N > 1$ we instead refer to its ring-variant, RLWE. While GLWE is defined over the torus $\mathbb T$, note that the integers modulo $q$ can be embedded into $\mathbb T$ with $\mathbb Z_q \cong \frac{1}{q} \mathbb Z / \mathbb Z \subset \mathbb T$. We thus depart from the torus notation and instead use $\mathbb Z_q$.

The torus-based FHE scheme TFHE~\cite{TFHE_original_paper} encrypts messages by encoding them into the noise of an GLWE sample. A message $m \in \mathfrak R_p^N$ is encrypted under the secret key $\sk \in \mathbb B_N[X]^k$ as follows:
\[
(\vec a, b = \langle \vec a, \sk \rangle + \Delta m + e),
\]
where $\vec a \sample \mathfrak R_{q, N}^k$ and all coefficients of $e$ are sampled from $\chi_\sigma$. By scaling $m$ with a large enough $\Delta$, the message is shifted into the MSBs and remains unaffected by the growing noise $e$, as long as $e$ remains small. We refer to $\vec a$ as the \emph{mask} of the ciphertext, and to $b$ as the \emph{body}. We denote with $\text{GLWE}_{\sk, \sigma}(\Delta m)$ the distribution of possible encryptions of $m$, scaled by $\Delta$, under the secret key $\sk$ and with noise distribution $\chi_\sigma$. 



TFHE~\cite{TFHE_original_paper} leverages LWE ciphertexts to encrypt integers. It provides addition, multiplication, as well as programmable bootstrapping (PBS). This differs from traditional bootstrapping by allowing the evaluation of an arbitrary lookup table for free during bootstrapping. RLWE ciphertexts are only applied during the computation of the PBS operation.

\smallsubsec{Parameters} TFHE uses both LWE and RLWE. We use $n$ to denote the dimension of the LWE ciphertext mask. Similarly, for RLWE ciphertexts $k$ refers to the mask dimension and $N$ to the polynomial degree. We denote with $p$ the plaintext modulus and with $q$ the ciphertext modulus. The scaling factor $\Delta$ is then given by $\Delta = \frac{q}{2p}$. We denote with $\beta$ the decomposition base, while $\ell$ denotes the decomposition depth.

\smallsubsec{Ciphertext Types} Messages in TFHE are encrypted as LWE ciphertexts. As noted above, RLWE ciphertexts are used during the bootstrapping operation. TFHE additionally requires GLev and GGSW ciphertexts. We informally describe the distributions of GLev and GGSW encrypting a message $m$. Note that, analogously to GLWE, GLev generalizes Lev and ring-Lev ciphertexts, and GGSW accordingly GSW and ring GSW ciphertexts.

\begin{align*}
&\text{GLev}_{\sk, \sigma}^{\beta, \ell}(m) = \left( GLWE_{\sk, \sigma}\left(\frac{q}{\beta^1}m\right), \dots, GLWE_{\sk, \sigma}\left(\frac{q}{\beta^\ell}m\right) \right)\\
&\text{GGSW}_{\sk, \sigma}^{\sk', \beta, \ell}(m) = \\
&\left( 
GLev^{\beta, \ell}_{\sk, \sigma}\left(m\cdot \sk'_1\right), 
\dots, 
GLev^{\beta, \ell}_{\sk, \sigma}\left(m\cdot \sk'_k\right), 
GLev^{\beta, \ell}_{\sk, \sigma}\left(m\right) 
\right)
\end{align*}


\smallsubsec{Key Generation} TFHE requires both an LWE key $\sk \sample \mathbb B^n$ and an RLWE key $\Sk \sample \mathbb B_N[X]^k$ which are sampled uniformly at random. Note that the RLWE secret key $\Sk$ can be represented as an LWE secret key denoted $\Sk'$. This is accomplished by simply concatenating the coefficient vectors of all polynomials in $\Sk$ into a single binary vector. This creates an LWE binary key of length $k \cdot N$.

TFHE additionally requires a bootstrapping key $\bsk$ and a key-switching key $\ksk$. The bootstrapping key consists of $k$ RGSW ciphertexts, each encrypting a bit $\sk_i$ under the RLWE key $\Sk$. Similarly, the key-switching key $\ksk$ consists of $k \cdot N$ Lev ciphertexts, each encrypting a bit of $\Sk'$ under $\sk$.
\[
\bsk = \left(\hatt C_i \in \rgsw_\Sk\left(\sk_i\right) \right)_{i=0}^{n-1}
\]
\[
\ksk = \left(\hat C_i \in \lev_\sk\left(\Sk'_i\right) \right)_{i=0}^{kN-1}
\]

\smallsubsec{Encryption} Given a scaled message $\Delta m$, TFHE applies the LWE encryption under $\sk$: $(\vec a, b = \langle \vec a, \sk \rangle + \Delta m + e) \leftarrow \text{LWE}_{\sk, \sigma}(\Delta m)$. 

\smallsubsec{Decryption} A GLWE ciphertext $c = (\vec a, b)$, is decrypted by computing $m = \left\lfloor \frac{b - \langle \vec a, \sk \rangle}{\Delta}\right\rceil$. Note that $b - \langle \vec a, \sk\rangle = \Delta m + e$ and thus $\left\lfloor \frac{\Delta m + e}{\Delta}\right\rceil = m$ as long as $||e||_\infty < \frac{\Delta}{2} = e_{max}$. This is the condition required for correct decryption.

\smallsubsec{Decomposition} A decomposition algorithm with quality $\beta$ and precision $\epsilon$ splits a value $x$ into smaller elements $(x_1, \dots, x_\ell)$ such that $||x_i||_\infty \leq \beta$ for $1 \leq i \leq \ell$ and $||x - \tilde x||_\infty \leq \epsilon$, where $\tilde x$ denotes the reconstructed value. The \emph{canonical decomposition} splits $x$ such that $\tilde x = \sum_{i=1}^\ell x_i\frac{q}{\beta^i}$. These concepts are defined both for scalars and polynomials. Additionally, the decomposition is applied element-wise when applied to a ciphertext.

\smallsubsec{Addition} Addition is native to GLWE ciphertexts. Two ciphertexts $(\vec a_1, b_1)$ and $(\vec a_2, b_2)$ respectively encrypting $m_1$ and $m_2$ can be added to form a ciphertext $(\vec a_1 + \vec a_2, b_1 + b_2)$ encrypting $m_1$ + $m_2$. The noise grows linearly with each addition.

\smallsubsec{External Product} Multiplication in TFHE is implemented via the external product. The algorithm takes a GGSW ciphertext $\hat{\hat{C}} \in \text{GGSW}(m_1)$ and a GLWE ciphertext $C = (\vec a, b) \in \text{GLWE}(m_2)$. The product is computed by
\[
C \boxdot \hat{\hat{C}} = \sum_{i=0}^{k+1}\sum_{j=1}^\ell \decomp(C)_{i,j} \cdot \hat{\hat{C}}_{i,j},
\label{eq:external_product}
\]
where $\decomp(C)_{i,j}$ refers to the $j$-th decomposition element of the $i$-th entry in $C$. The external product results in a GLWE ciphertext encrypting $m_1 \cdot m_2$. Note that the noise of the resulting ciphertext is considerably higher than the noise in $\hat{\hat{C}}$ and in $C$.

\smallsubsec{CMUX} The CMUX, or controlled mutex, operation takes two GLWE ciphertexts $c_0$ and $c_1$ respectively encrypting $m_0$ and $m_1$, as well as a GGSW ciphertext $\hat{\hat C}$ encrypting a single bit $b$, and computes
\[
\oper{CMUX}(\hat{\hat C}, c_0, c_1) = c_0 - (c_0 + c_1) \boxdot \hat{\hat C}.
\]
This results in a GLWE ciphertext encrypting $m_b$.

\smallsubsec{Programmable Bootstrap}\label{sec:pbs} The programmable bootstrapping operation consists of a blind rotation followed by a sample extract. Let $acc$ be an RLWE ciphertext encrypting some polynomial $F$. Let $c = (\vec a, b)$ be the LWE ciphertext to be bootstrapped, and $\bsk$ the bootstrapping key. The blind rotation is performed by first setting $acc = acc \cdot X^{-b}$, and then iteratively computing $$acc = \oper{CMUX}(\bsk_i, acc, acc\cdot X^{a_i}),$$
for each $a_i \in \vec a$. This results in an RLWE encryption of $F \cdot X^{\langle \vec a, \sk \rangle - b}$, effectively rotating the $\Delta m + e$-th coefficient of $F$ to the $0$-th position.

Let $LUT$ be a lookup table. Consider a polynomial $F$ constructed as $F(X) = \sum_{m \in \mathbb Z_p}\sum_{|e| < e_{max}} LUT[m] \cdot X^{\Delta m + e}$. Blindly rotating an encryption of $F$ effectively rotates a coefficient holding the value $LUT[m]$ to the $0$-th position.
This coefficient can be extracted into an LWE ciphertext via sample extraction. Let $d_{i,j}$ be the $j$ coefficient of the $i$-th polynomial in an RLWE ciphertext. For a given index $h$ compute an LWE ciphertext $(\vec a', b')$ with $kN$ mask elements:
\[
    b' = d_{k, h}  \hspace{0.5cm}\text{~~~~ and ~~~~}\hspace{0.5cm}
    a'_{iN+j} = \begin{cases}
        d_{i, h-j} &\text{if } 0 \leq j \leq h,\\
        -d_{i, h-j} &\text{otherwise.}
    \end{cases}
\]
The ciphertext $(\vec a', b')$ encrypts the $h$-th coefficient of the given RLWE ciphertext.
In summary, the programmable bootstrapping blindly rotates the encrypted polynomial $F$, and subsequently extracts the $0$-th coefficient via the sample extract. Note that the result is encrypted under $\Sk'$, and thus has mask dimension $kN$.

\smallsubsec{Key Switch}
The key-switching operation allows a ciphertext encrypted under one key to be modified to encrypt the same message under a different key without decryption. Concretely, the key-switching key $\ksk$ allows an LWE ciphertext $c = (\vec a, b)$ encrypted under $\Sk'$ to be changed to an encryption under $\sk$. This is accomplished by computing
\[
(0, \dots, 0, b) - \sum_{i=0}^{kN - 1}\sum_{j=1}^\ell\decomp(a_i)_j, \ksk_{i,j}.
\]
The key-switching operation is applied after a programmable bootstrapping in order to return bootstrapped ciphertexts to encryptions under the original LWE key $\sk$.

\subsection{Number Theoretic Transform}
The discrete Fourier transform (DFT)~\cite{ntt_guide} enables efficient polynomial multiplication. Polynomials represented in the Fourier domain can be multiplied by simple element-wise multiplication, as opposed to the convolution of coefficients otherwise required. Let $g$ be a polynomial with coefficients $g_0, \dots, g_{N-1}$. An $N$-point DFT is defined as the evaluation of $g$ at the $N$-th roots of unity generated by the primitive $N$-th root of unity $\omega$. That is
$$\dft_N(g) = (\hat g_0, \dots, \hat g_{N-1}), \text{ with } \hat g_k = g(\omega^k),$$ where $\omega^N = 1$ and $\omega^k \neq 1$ for $1 \leq k < N$. In order to compute the inverse DFT ($\idft$) the polynomial $\hat g$ with coefficients $\hat g_0, \dots, \hat g_{N-1}$ is evaluated at the inverted roots of unity $\omega^{-0}, \dots, \omega^{-(N-1)}$ and scaled by $N^{-1}$. In summary the following holds: 
$$\hat g_k = \sum_{i=0}^{N-1}g_i\omega^{ki}  \hspace{0.5cm}\text{  and  }\hspace{0.5cm}  g_k = \frac{1}{N} \sum_{i=0}^{N-1}\hat g_i\omega^{-ik}$$


When working over real-valued or complex-valued polynomials, the existence of a primitive $N$-th root of unity is guaranteed and given by $\omega = \exp(\frac{2i\pi}{N})$. In a finite field $\oper{GF}(p)$ such a root of unity exists if and only if $N$ divides $p - 1$. Finding a primitive root of unity in a given finite field is not trivial.


\smallsubsec{Cyclicy, Negacyclicy and Twisting}
Note that in order to accurately represent the product of two polynomials of degree $<N$, a $(2N-1)$-point DFT is necessary. Thus $g = \idft_{2N-1}(\dft_{2N-1}(p) * \dft_{2N-1}(q))$. If instead $g$ is computed by leveraging an $N$-point DFT, a positive-wrapped cyclic reduction is performed on the product $g$. That is, the polynomial $g$ is reduced modulo $X^N - 1$, implying the congruence $X^N \equiv 1$ and thus
$$
g(x) = \sum_{i=0}^{2N -1} g_ix^i = \sum_{i=0}^{N-1} (g_i + g_{N+i})x^i \bmod (X^N - 1).
$$.

When working over the cyclotomic polynomial ring $\mathfrak R_{q, N} = \mathbb Z_q[X]/(X^N + 1)$, as needed for TFHE, a \emph{negacyclic} reduction is required. That is, $g$ should be reduced modulo $X^N + 1$, implying the congruence $X^N \equiv -1$. In this case, the following holds:
$$
g(x) = \sum_{i=0}^{2N -1} g_ix^i = \sum_{i=0}^{N-1} (g_i - g_{N+i})x^i \bmod (X^N + 1).
$$

The DFT can be leveraged to perform a negacyclicly reduced polynomial multiplication, by 'twisting` the roots of unity before evaluating the DFT. In order to twist an $N$-point DFT a $2N$-th primitive root of unity $\psi$ is required. The polynomial $g$ is then evaluated at the roots of unity $\omega^i\cdot \psi$ for $0 \leq i < N$. Note that $\psi^2 = \omega$, and thus $\omega^i\cdot \psi = \psi^{2i + 1}$. In the complex plane this can be visualized as rotating the $N$-th roots of unity $\omega^i$ by the $2N$-th primitive root of unity $\psi$ on the unit circle. $\psi$ hereby lies halfway between $\omega^0$ and $\omega^1$.

\smallsubsec{Number Theoretic Transform (NTT)}
Several algorithms exist to efficiently compute the DFT. These algorithms are denoted as fast Fourier transform (FFT), though this name is often used specifically when operating over real or complex-valued polynomials. When working over prime-order finite fields, the term number theoretic transform (NTT) is used. The cyclotomic FFT (CFFT) generalizes this to the case of general finite fields.

The most common DFT algorithms rely on the symmetric properties of roots of unity, as well as the fact that an $N$-point DFT can be expressed in terms of $N_1$ separate $N_2$-point DFTs, where $N = N_1\cdot N_2$. Consider concretely a power-of-two $N$. The following holds for a polynomial $g$ with coefficients $g_0, \dots, g_{N-1}$, and its DFT $\hat g$:
\begin{align*}
    \hat g_k &= \sum_{i = 0}^{N-1}g_i\omega^{ik}\\
    &= \sum_{i = 0}^{N/2-1}g_{2i}\omega^{2ik} + \sum_{i = 0}^{N/2-1}g_{2i +1 }\omega^{(2i + 1)k}\\
    &= \sum_{i = 0}^{N/2-1}g_{2i}\omega^{2ik} + \omega^k \cdot \sum_{i = 0}^{N/2-1}g_{2i +1 }\omega^{2ik}\\
    &= \hat E_k + \omega^k \hat O_k,
\end{align*}
where $\hat E$ and $\hat O$ are respectively the $N/2$-point DFTs of the even- and odd-indexed coefficients of $g$. Note additionally that $\omega^{k + N/2} = -\omega^k$ and thus $\hat g_{k + N/2} = \hat E_k - \omega^k \hat O_k$. 

The same symmetries can be leveraged for the efficient computation of negacyclic DFTs. In this case, the following condition holds:
\begin{align*}
    \hat g_k &= \sum_{i = 0}^{N-1}g_i\psi^{(2k+1) \cdot i}\\
    &= \hat E_k + \psi^{2k + 1} \hat O_k,\\
    \hat g_{k + N/2} &= \hat E_k - \psi^{2k + 1} \hat O_k.
\end{align*} 

\begin{figure}
    \centering

        \begin{subfigure}{.5\linewidth}
          \centering
         \includegraphics[width=0.8\columnwidth]{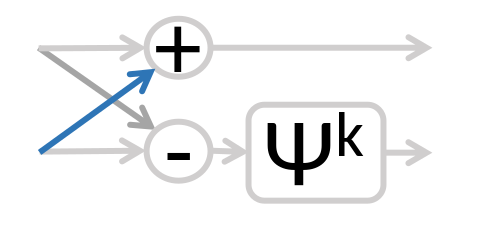}
          \vspace{-0.1in}
          \caption{Gentleman-Sande}
          \label{fig:gentleman}
        \end{subfigure}%
        \begin{subfigure}{.5\linewidth}
          \centering   
         \includegraphics[width=0.8\columnwidth]{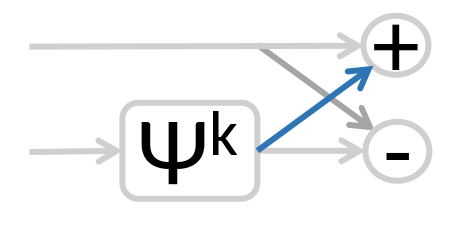}
          \vspace{-0.1in}
          \caption{Cooley-Tukey}
          \label{fig:cooley}
        \end{subfigure}
    
    \caption{Butterfly Configurations}
    \label{fig:butterfly_configurations}
\end{figure}

\begin{figure}
    \centering
    \includegraphics[width=\columnwidth]{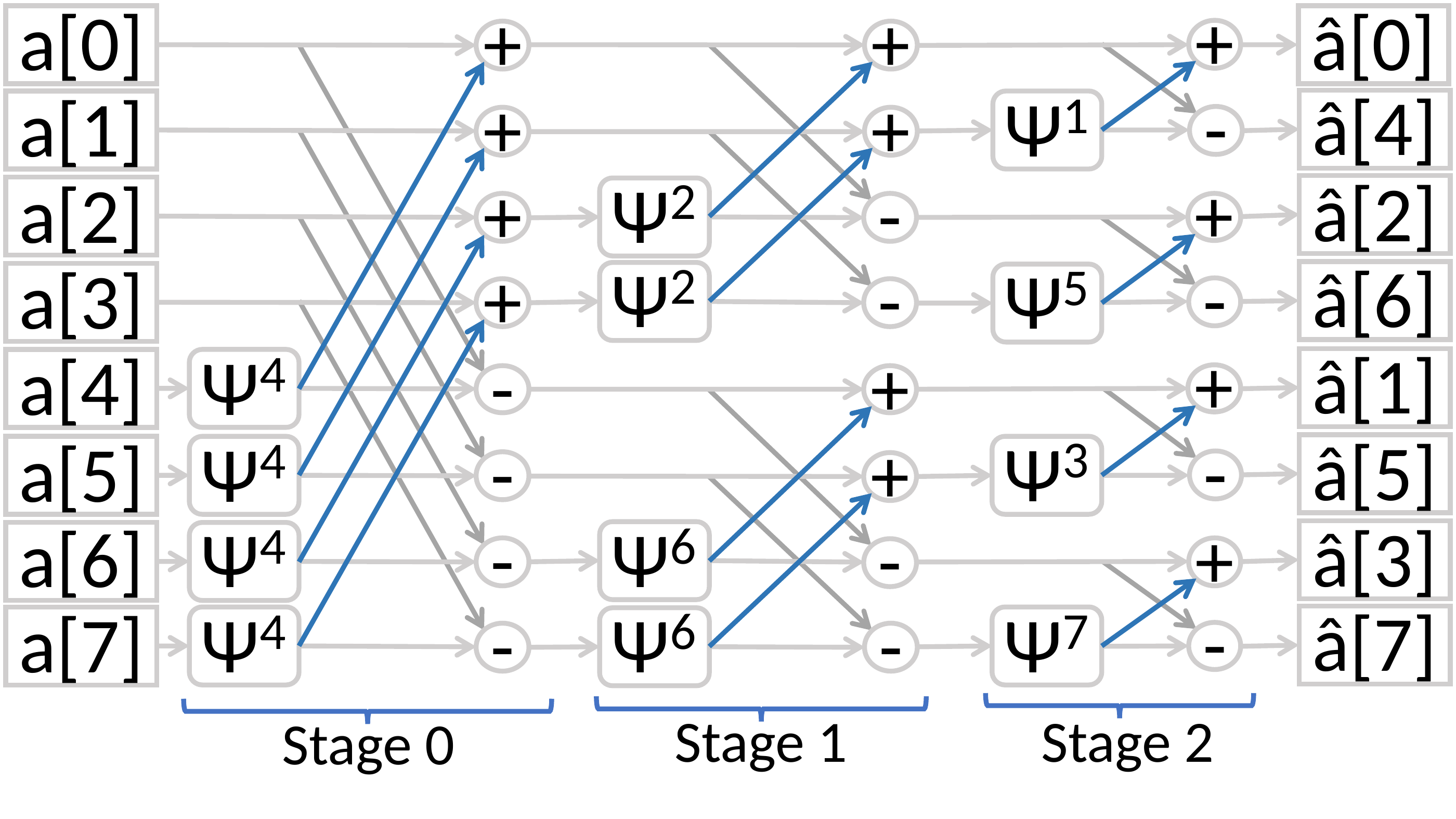}
    \caption{8-point Cooley-Tukey NTT with normal-ordered input and bit-reversed-ordered output.}
    \label{fig:ct-8-point-ntt}
\end{figure}

These symmetries allow the implementation of Cooley-Tukey (CT) butterfly construction as illustrated in Figure~\ref{fig:cooley}. A CT-butterfly takes a root of unity $\psi^k$ as well as two values $a_0, a_1$ and computes the outputs $a_0' = a_0 + \psi^k \cdot a_1$ and $a'_1 = a_0 - \psi^k \cdot a_1$. For $k = 1$, this computes a $2$-point NTT. From this construction, an $N$-point NTT can be built recursively. A construction for a negacylcic $8$-point is depicted in figure~\ref{fig:ct-8-point-ntt}. We denote each recursive step as a stage. Starting at $0$, each stage $i$ performs $(i + 1) \times \frac{N}{2^i}$ butterflies.

Note that the output of this process is not ordered as expected. Instead, the $N$-point Cooley-Tukey NTT returns values in bit-reversed order. The expected value $\hat a_i$ is placed in the $\bar i$-th position of the output, where $\bar i$ is the value $i$ such that its $\log N$-bit representation is reversed. The value $\hat a_4$ of an $8$-point NTT for example is placed at index $\overline{4} = \overline{100}_2 = 001_2 = 1$.

\smallsubsec{Inverse NTT}
The inverse NTT (iNTT) can be similarly computed recursively. Indeed, an $N$-point Cooley-Tukey NTT can simply be reversed to compute the inverse NTT. For this, at each point the roots of unity $\psi^k$ are multiplicatively inverted to $\psi^{-k}$. However, the output of an $N$-point CT NTT is not compatible with the input of an $N$-point CT iNTT. This is due to the bit-reversed order of the outputs.

\begin{figure}
    \centering
    \includegraphics[width=\columnwidth]{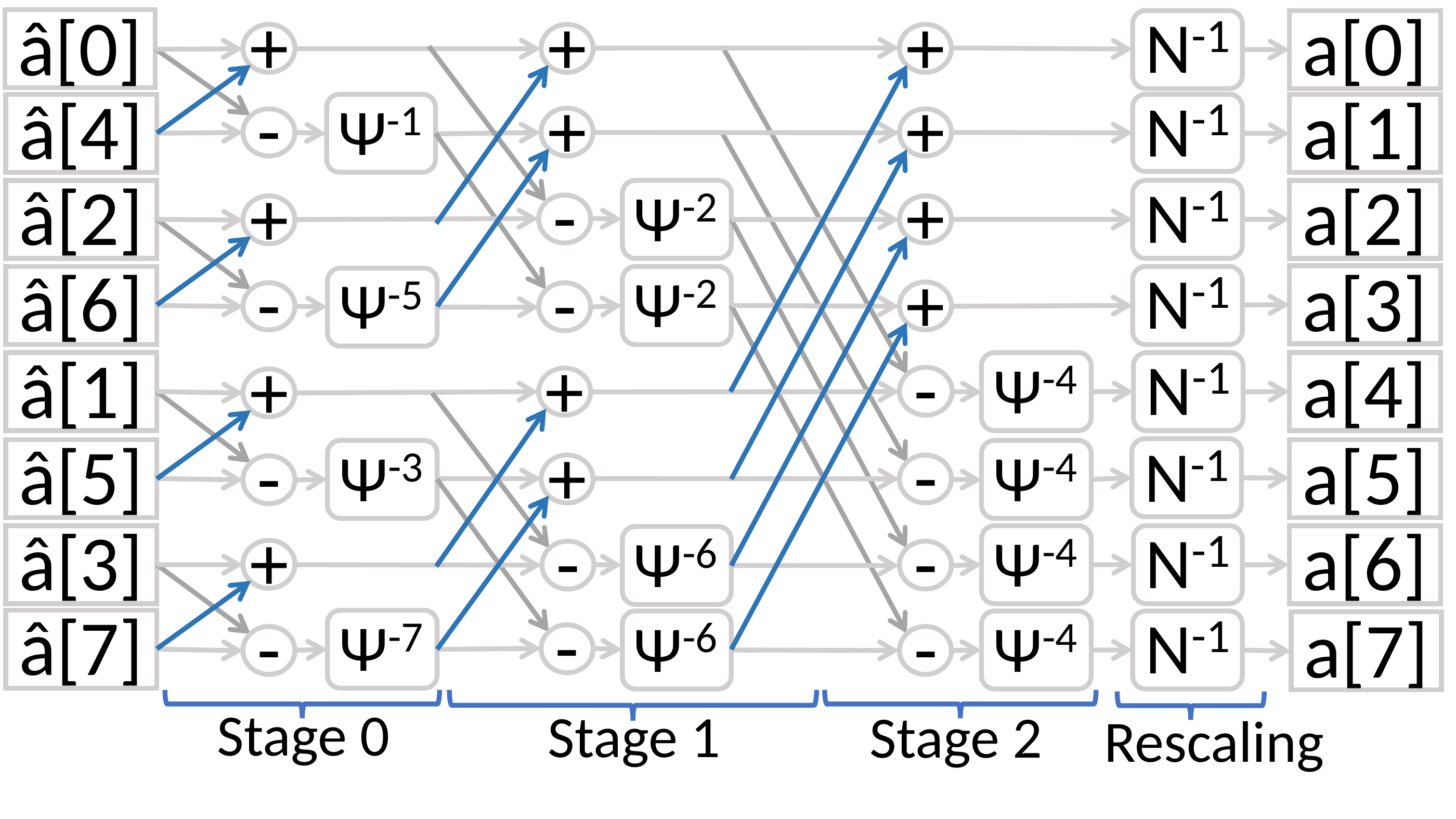}
    \caption{8-point Gentleman-Sande iNTT with bit-reversed-ordered input and normal-ordered output.}
    \label{fig:gs-8-point-intt}
\end{figure}

Instead, the Gentleman-Sande (GS) butterfly construction can be leveraged to avoid expensive reordering of the inputs as illustrated in Figure~\ref{fig:gentleman}. On two inputs $\hat a_0$ and $\hat a_1$, as well as the root of unity $\psi^k$, the GS butterfly computes the outputs $\hat a'_0 = \hat a_0 + \hat a_1$ and $\hat a'_1 = \hat a_0 - \psi^{-k}\cdot \hat a_1$. This computes the $2$-point iNTT up to a scaling by $2^{-1}$. GS butterflies can be leveraged to construct the iNTT. This is accomplished by reversing the order of the stages in the $N$-point CT NTT construction, replacing each CT butterfly with a GS butterfly, and multiplicatively inverting the corresponding roots of unity. The output of the last stage must additionally be scaled by $N^{-1}$ to complete the iNTT. Note that, given reversed-bit-ordered inputs, this construction yields normal-ordered outputs, as illustrated by Figure~\ref{fig:gs-8-point-intt}.

\subsection{Field Programmable Gate Arrays} \label{sec:fpga}
Field-Programmable Gate Arrays (FPGAs)~\cite{fpga_vs_asic_basics} are programmable integrated circuits composed of arithmetic and logic units, memory blocks, and routing interconnects. They offer high computational performance and energy efficiency compared to general-purpose hardware such as CPUs and GPUs, while remaining far more flexible and cost-effective than ASICs, whose development is expensive and time-consuming~\cite{jayasena2024directed}.

FPGAs are also well-suited for hardware prototyping and testing, as their reprogrammable nature allows rapid iteration and validation of hardware designs before committing to costly ASIC fabrication~\cite{jayasena2024directed}.
An FPGA consists mainly of digital signal processing (DSP) units for arithmetic, lookup tables (LUTs) for logic, flip-flops (FFs) for bit storage, and various on-chip RAMs. Designs are written in a hardware description language (HDL) and synthesized into routing configurations that define signal paths between elements.
FPGA performance depends on the operating frequency, which is constrained by signal propagation delays across routed connections. Therefore, efficient design requires minimizing long interconnects and balancing resource utilization to achieve high clock speeds and reliable operation.

\section{Implementation Methodology}\label{sec:implementation}

In this section, we present the methodology followed in implementing our TFHE processor. Specifically, we describe the optimization strategies applied to the modular reduction, NTT, external product, blind rotation, programmable bootstrapping, key switching, and MulAdd modules in order to outperform the state of the art.

\subsection{Modular Reduction}\label{sec:modred}

Modular reduction ensures that values propagated through the design remain within $\mathbb{Z}_q$, preventing uncontrolled growth during computations. We implement two types of reductions depending on the operation and optimize both through modulus selection and algorithmic efficiency.

\smallsubsec{Reduction Types}
When multiplying values in $\mathbb{Z}_q$, a full reduction modulo $q$ is required to constrain results to the interval $[0, q)$. We exploit the structure of the chosen Solinas prime to efficiently implement this \emph{full reduction}.
For additions or subtractions in $\mathbb{Z}_q$, values may only exceed or fall below $[0, q)$ by at most $q-1$. In such cases, a single conditional addition or subtraction of $q$ suffices to restore the result to $\mathbb{Z}_q$. We refer to this as a \emph{simple reduction}.

\smallsubsec{Efficient Reduction Algorithms}
A naive reduction by division is too costly for hardware implementation. The Barrett reduction~\cite{barrett} offers an efficient alternative for arbitrary moduli and can be combined with Montgomery multiplication for modular products. However, both methods still require complex arithmetic and are less suited to high-throughput hardware.
Instead, we leverage moduli that simplify reduction. Power-of-two moduli allow implicit reduction via overflow, while Mersenne and Solinas primes enable reduction using only shifts and additions. Such moduli significantly reduce hardware complexity and latency.

\smallsubsec{Choice of Modulus}
We adopt the NTT-friendly 64-bit Solinas prime:
\[
q = 2^{64} - 2^{32} + 1.
\]
This prime offers efficient modular reduction using only additions and bit shifts, while remaining compatible with standard TFHE parameter sets. It also supports suitable primitive roots of unity required for NTT operations.

Given $2^{96} \equiv -1 \pmod q$ and $2^{64} \equiv 2^{32} - 1 \pmod q$, a 128-bit intermediate product $v$ can be reduced as:
\[
\begin{array}{lllll}
v &= a \cdot 2^{96} &+ b \cdot 2^{64} &+ c \cdot 2^{32} &+ d \\[4pt]
  &= a(-1) &+ b(2^{32} - 1) &+ c \cdot 2^{32} &+ d \\[4pt]
  &\multicolumn{4}{l}{
  = (b + c) \cdot 2^{32} + (-a - b + d) \bmod q.}
\end{array}
\]
The value $v$ is split into four 32-bit segments, followed by three additions and one shift to compute the reduced result. A final conditional addition ensures the value lies in $\mathbb{Z}_q$.

\smallsubsec{Scalar Multiplication}
We employ the Karatsuba algorithm to accelerate scalar multiplication. For two $\ell$-bit operands:
\begin{align*} x\cdot y &= (x_1 \cdot 2^{\ell/2} + x_2) \cdot (y_1 \cdot 2^{\ell/2} + y_2) \\ &= x_1 \cdot y_1\cdot 2^{\ell} + (x_1\cdot y_0 + x_0 \cdot y_1) \cdot 2^{\ell/2} + x_2 \cdot y_2 \end{align*}
reducing the number of required multiplications from four to three at each recursion level. We apply a recursion depth of two, balancing logic resource usage and performance.

\subsection{Number Theoretic Transform (NTT)}\label{subsec:ntt}

The Number Theoretic Transform (NTT) combines the efficiency of FFT-based polynomial multiplication with exact finite-field arithmetic, eliminating rounding errors and simplifying noise management in TFHE. By selecting an appropriate prime modulus that supports the necessary roots of unity, the NTT enables precise, high-throughput negacyclic multiplications fully compatible with TFHE parameters and well-suited for FPGA implementation.

Building on these advantages, we apply the NTT to accelerate the most computationally intensive component in TFHE: the negacyclic polynomial multiplication. Its cost dominates overall performance due to the large polynomial sizes and the frequency of these operations. For instance, a single programmable bootstrapping (PBS) requires $n \times (k + 1) \times (k + 1) \times \ell$ negacyclic multiplications. To mitigate this bottleneck, we employ an NTT-based multiplier using the Solinas prime $q = 2^{64} - 2^{32} + 1$, which both accelerates modular reduction and enables efficient negacyclic convolution.


\smallsubsec{Butterfly Construction} 
Our construction starts with a single butterfly. Given two values, $a_i$ and $a_j$, as well as a twiddle factor $\psi^k$, the butterfly computes $a_i + \psi^k \cdot a_j$ and $a_i - \psi^k \cdot a_j$. That is, a butterfly computes a 2-point NTT with the given root of unity $\psi^k$. We instantiate several such butterflies into a \emph{butterfly block} (BF-block) as we discussed in Section~\ref{sec:tfhe}. For a given throughput $\throu$ a butterfly block contains $\frac{\throu}{2}$ butterflies, allowing $\throu$ entries to be processed at once. Each BF-block has fixed-routing paths. We instantiate a single BF-block for each stage of the NTT. Per NTT, each stage processes $N$ inputs in groups of size $\throu$. How inputs are paired and in what order they are passed into the stage's BF-block is handled individually per stage. Additionally, depending on the current input being processed, different twiddle factors need to be used. To solve this, each butterfly has a dedicated memory resource from which it reads the precomputed twiddle factors sequentially. 

\smallsubsec{Forward and Inverse Transform Architecture}
Using this method, we decompose an $N$-point NTT into smaller sub-transforms until reaching the stage that computes $\throu$-point NTTs. At this point, the architecture transitions to a fully parallel configuration, implementing a complete $\throu$-point NTT as the final stage without intermediate buffers. This design ensures a sustained throughput of $\throu$ while minimizing latency.

\begin{figure}[htp]
    \centering
    {\tiny\input{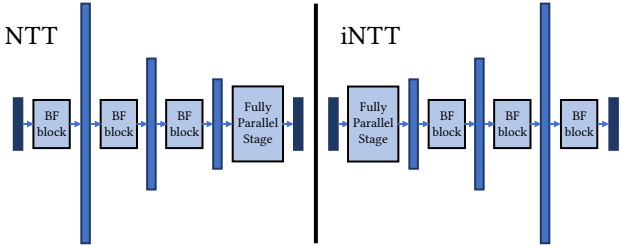}}
    \vspace{-0.1in}
    \caption{Structure of the NTT and iNTT for the case $\mathcal{T} = \log N - 3$. Blue bars represent stage input buffers; their height indicates relative size.}
    \label{fig:our_ntt_structure}
\end{figure}

The inverse NTT (iNTT) mirrors the forward transform, employing Gentleman–Sande butterflies and multiplicatively inverted twiddle factors. Outputs are naturally produced in normal order, removing the need for bit-reversal permutations or complex reordering networks. The overall structure is illustrated in Figure~\ref{fig:our_ntt_structure}.

\smallsubsec{Buffering}
Each stage in the NTT depends on the output of the previous one. Since every stage processes $N$ values, a naive implementation would require $N \log N$ buffers, or twice that when pipelined, as each stage would need dedicated input and output buffers to avoid overwriting intermediate data.
We minimize this requirement through careful scheduling. The $i$-th stage performs an $N/2^i$-point NTT $(i+1)$ times, where each computation is independent. By ensuring that all inputs for one such NTT are supplied before starting the next, each stage only needs an output buffer of size $N/2^i$, halving buffer size at every level.

\smallsubsec{Inter-Stage Data Reuse} Moreover, we reuse memory between stages: outputs overwrite inputs of subsequent stages that have already been consumed.
Further optimization is achieved by directly streaming certain outputs to the next stage, bypassing buffers entirely. When a stage completes processing and the next begins, a subset of outputs can be passed forward immediately. Dataflow analysis shows that one-quarter of all values can bypass buffering through this strategy.

Overall our NTT implementation approach requires $\log(N/\throu)$ stages, each with an output buffer of size $\frac{3}{4}\cdot\frac{N}{2^i}$. This leads to a total of $\frac{3}{4}N \cdot \sum_{i=0}^{\log(N/\throu)}2^{-i}$ values per NTT.

\subsection{External Product}\label{sec:extprod}
As we discussed in Section~\ref{sec:tfhe}, it takes two inputs: an RLWE $C$ ciphertext and an RGSW $\hatt C$ ciphertext for the external product implementation. The RLWE ciphertext is first decomposed, then all entries in the decomposed RLWE ciphertext are multiplied element-wise with all elements in the RGSW ciphertext. All resulting products are added together, resulting in the desired external product:
\[
\oper{ExtProd}(C, \hatt C ) = \sum_{i=0}^{k}\sum_{j=1}^\ell \decomp(C)_{i,j} \cdot \hatt C_{i,j}.
\]
Note that $\decomp(C)_{i,j}$ refers to the $j$-th decomposition element of the $i$-th entry in $rlwe$. This element is a polynomial. $\hatt C_{i,j}$ refers to an RLWE ciphertext in $\hatt C$. Thus the product $\decomp(C)_{i,j} \cdot \hatt C_{i,j}$ consists of $k + 1$ polynomial multiplications.

We construct an external product module in five parts, dedicated to the decomposition, the number theoretic transform, the element-wise multiplication of transformed polynomials, the addition of products, and finally the computation of the inverse NTT. 

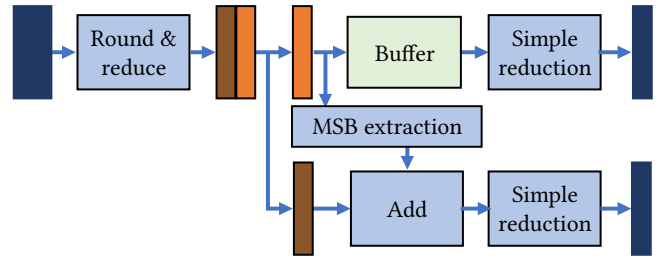
\begin{figure}[htp]
    \centering
    \tikzset{every picture/.style={line width=0.75pt}} 

\begin{tikzpicture}[x=0.75pt,y=0.75pt,yscale=-1,xscale=1]

\draw [color={rgb, 255:red, 68; green, 114; blue, 196 }  ,draw opacity=1 ][line width=1.5]    (240.91,42.29) -- (252.52,42.29) ;
\draw [shift={(256.52,42.29)}, rotate = 180] [fill={rgb, 255:red, 68; green, 114; blue, 196 }  ,fill opacity=1 ][line width=0.08]  [draw opacity=0] (6.97,-3.35) -- (0,0) -- (6.97,3.35) -- cycle    ;
\draw [color={rgb, 255:red, 68; green, 114; blue, 196 }  ,draw opacity=1 ][line width=1.5]    (35.5,42.29) -- (44.33,42.29) ;
\draw [shift={(48.33,42.29)}, rotate = 180] [fill={rgb, 255:red, 68; green, 114; blue, 196 }  ,fill opacity=1 ][line width=0.08]  [draw opacity=0] (6.97,-3.35) -- (0,0) -- (6.97,3.35) -- cycle    ;
\draw  [fill={rgb, 255:red, 180; green, 199; blue, 231 }  ,fill opacity=1 ] (49.17,23.54) -- (105.65,23.54) -- (105.65,61.05) -- (49.17,61.05) -- cycle ;

\draw  [color={rgb, 255:red, 32; green, 56; blue, 100 }  ,draw opacity=1 ][fill={rgb, 255:red, 32; green, 56; blue, 100 }  ,fill opacity=1 ] (17,19) -- (36,19) -- (36,65.58) -- (17,65.58) -- cycle ;
\draw  [fill={rgb, 255:red, 226; green, 240; blue, 217 }  ,fill opacity=1 ] (185.39,24.39) -- (242.53,24.39) -- (242.53,60.19) -- (185.39,60.19) -- cycle ;

\draw [color={rgb, 255:red, 68; green, 114; blue, 196 }  ,draw opacity=1 ][line width=1.5]    (215.53,88.88) -- (215.53,97.59) ;
\draw [shift={(215.53,101.59)}, rotate = 270] [fill={rgb, 255:red, 68; green, 114; blue, 196 }  ,fill opacity=1 ][line width=0.08]  [draw opacity=0] (6.97,-3.35) -- (0,0) -- (6.97,3.35) -- cycle    ;
\draw [color={rgb, 255:red, 68; green, 114; blue, 196 }  ,draw opacity=1 ][line width=1.5]    (174.22,42.29) -- (174.22,66.26) ;
\draw [shift={(174.22,70.26)}, rotate = 270] [fill={rgb, 255:red, 68; green, 114; blue, 196 }  ,fill opacity=1 ][line width=0.08]  [draw opacity=0] (6.97,-3.35) -- (0,0) -- (6.97,3.35) -- cycle    ;
\draw [color={rgb, 255:red, 68; green, 114; blue, 196 }  ,draw opacity=1 ][line width=1.5]    (131.44,42.29) -- (153.5,42.29) ;
\draw [shift={(157.5,42.29)}, rotate = 180] [fill={rgb, 255:red, 68; green, 114; blue, 196 }  ,fill opacity=1 ][line width=0.08]  [draw opacity=0] (6.97,-3.35) -- (0,0) -- (6.97,3.35) -- cycle    ;
\draw [color={rgb, 255:red, 68; green, 114; blue, 196 }  ,draw opacity=1 ][line width=1.5]    (167.11,121.21) -- (183,121.21) ;
\draw [shift={(187,121.21)}, rotate = 180] [fill={rgb, 255:red, 68; green, 114; blue, 196 }  ,fill opacity=1 ][line width=0.08]  [draw opacity=0] (6.97,-3.35) -- (0,0) -- (6.97,3.35) -- cycle    ;
\draw [color={rgb, 255:red, 68; green, 114; blue, 196 }  ,draw opacity=1 ][line width=1.5]    (145.22,42.29) -- (145.22,121.21) -- (153.8,121.21) ;
\draw [shift={(157.8,121.21)}, rotate = 180] [fill={rgb, 255:red, 68; green, 114; blue, 196 }  ,fill opacity=1 ][line width=0.08]  [draw opacity=0] (6.97,-3.35) -- (0,0) -- (6.97,3.35) -- cycle    ;
\draw [color={rgb, 255:red, 68; green, 114; blue, 196 }  ,draw opacity=1 ][line width=1.5]    (168.33,42.29) -- (181,42.29) ;
\draw [shift={(185,42.29)}, rotate = 180] [fill={rgb, 255:red, 68; green, 114; blue, 196 }  ,fill opacity=1 ][line width=0.08]  [draw opacity=0] (6.97,-3.35) -- (0,0) -- (6.97,3.35) -- cycle    ;
\draw  [fill={rgb, 255:red, 180; green, 199; blue, 231 }  ,fill opacity=1 ] (256.52,23.54) -- (313,23.54) -- (313,61.05) -- (256.52,61.05) -- cycle ;

\draw  [fill={rgb, 255:red, 180; green, 199; blue, 231 }  ,fill opacity=1 ] (256.52,102.46) -- (313,102.46) -- (313,139.96) -- (256.52,139.96) -- cycle ;

\draw  [fill={rgb, 255:red, 180; green, 199; blue, 231 }  ,fill opacity=1 ] (187.07,102.46) -- (243.55,102.46) -- (243.55,139.96) -- (187.07,139.96) -- cycle ;

\draw  [fill={rgb, 255:red, 180; green, 199; blue, 231 }  ,fill opacity=1 ] (157.34,69.51) -- (252.99,69.51) -- (252.99,90) -- (157.34,90) -- cycle ;

\draw  [color={rgb, 255:red, 32; green, 56; blue, 100 }  ,draw opacity=1 ][fill={rgb, 255:red, 32; green, 56; blue, 100 }  ,fill opacity=1 ] (328.8,97.92) -- (338.3,97.92) -- (338.3,144.5) -- (328.8,144.5) -- cycle ;
\draw  [color={rgb, 255:red, 32; green, 56; blue, 100 }  ,draw opacity=1 ][fill={rgb, 255:red, 32; green, 56; blue, 100 }  ,fill opacity=1 ] (329.3,19) -- (338.8,19) -- (338.8,65.58) -- (329.3,65.58) -- cycle ;
\draw  [color={rgb, 255:red, 0; green, 0; blue, 0 }  ,draw opacity=1 ][fill={rgb, 255:red, 139; green, 87; blue, 42 }  ,fill opacity=1 ] (158.3,97.92) -- (167.8,97.92) -- (167.8,144.5) -- (158.3,144.5) -- cycle ;
\draw  [color={rgb, 255:red, 0; green, 0; blue, 0 }  ,draw opacity=1 ][fill={rgb, 255:red, 237; green, 125; blue, 49 }  ,fill opacity=1 ] (157.8,19) -- (167.3,19) -- (167.3,65.58) -- (157.8,65.58) -- cycle ;
\draw  [color={rgb, 255:red, 0; green, 0; blue, 0 }  ,draw opacity=1 ][fill={rgb, 255:red, 237; green, 125; blue, 49 }  ,fill opacity=1 ] (129.3,19) -- (138.8,19) -- (138.8,65.58) -- (129.3,65.58) -- cycle ;
\draw  [color={rgb, 255:red, 0; green, 0; blue, 0 }  ,draw opacity=1 ][fill={rgb, 255:red, 139; green, 87; blue, 42 }  ,fill opacity=1 ] (119.3,19) -- (128.8,19) -- (128.8,65.58) -- (119.3,65.58) -- cycle ;
\draw [color={rgb, 255:red, 68; green, 114; blue, 196 }  ,draw opacity=1 ][line width=1.5]    (106,42.29) -- (114.83,42.29) ;
\draw [shift={(118.83,42.29)}, rotate = 180] [fill={rgb, 255:red, 68; green, 114; blue, 196 }  ,fill opacity=1 ][line width=0.08]  [draw opacity=0] (6.97,-3.35) -- (0,0) -- (6.97,3.35) -- cycle    ;
\draw [color={rgb, 255:red, 68; green, 114; blue, 196 }  ,draw opacity=1 ][line width=1.5]    (311.91,42.29) -- (324.5,42.29) ;
\draw [shift={(328.5,42.29)}, rotate = 180] [fill={rgb, 255:red, 68; green, 114; blue, 196 }  ,fill opacity=1 ][line width=0.08]  [draw opacity=0] (6.97,-3.35) -- (0,0) -- (6.97,3.35) -- cycle    ;
\draw [color={rgb, 255:red, 68; green, 114; blue, 196 }  ,draw opacity=1 ][line width=1.5]    (242.41,121.21) -- (254.02,121.21) ;
\draw [shift={(258.02,121.21)}, rotate = 180] [fill={rgb, 255:red, 68; green, 114; blue, 196 }  ,fill opacity=1 ][line width=0.08]  [draw opacity=0] (6.97,-3.35) -- (0,0) -- (6.97,3.35) -- cycle    ;
\draw [color={rgb, 255:red, 68; green, 114; blue, 196 }  ,draw opacity=1 ][line width=1.5]    (312.91,121.21) -- (324.52,121.21) ;
\draw [shift={(328.52,121.21)}, rotate = 180] [fill={rgb, 255:red, 68; green, 114; blue, 196 }  ,fill opacity=1 ][line width=0.08]  [draw opacity=0] (6.97,-3.35) -- (0,0) -- (6.97,3.35) -- cycle    ;

\draw (77.63,42.29) node   [align=left] {\begin{minipage}[lt]{38.11pt}\setlength\topsep{0pt}
\begin{center}
Round \& reduce
\end{center}

\end{minipage}};
\draw (214.26,42.61) node   [align=left] {\begin{minipage}[lt]{39.27pt}\setlength\topsep{0pt}
\begin{center}
Buffer
\end{center}

\end{minipage}};
\draw (284.98,42.29) node   [align=left] {\begin{minipage}[lt]{38.11pt}\setlength\topsep{0pt}
\begin{center}
Simple reduction
\end{center}

\end{minipage}};
\draw (284.98,121.21) node   [align=left] {\begin{minipage}[lt]{38.11pt}\setlength\topsep{0pt}
\begin{center}
Simple reduction
\end{center}

\end{minipage}};
\draw (215.53,121.21) node   [align=left] {\begin{minipage}[lt]{38.11pt}\setlength\topsep{0pt}
\begin{center}
Add
\end{center}

\end{minipage}};
\draw (205.53,79.76) node   [align=left] {\begin{minipage}[lt]{64.54pt}\setlength\topsep{0pt}
\begin{center}
MSB extraction
\end{center}

\end{minipage}};

\end{tikzpicture}
    \caption{Structure of the decomposition module for $\ell = 2$.}
    \label{fig:decompose_module}
    \vspace{-0.1in}
\end{figure}

\smallsubsec{Preprocessing} The pipelined decomposition module decomposes coefficients in a polynomial by rounding and removing $\log q - \ell\cdot \log \beta$ bits. The remaining $\ell$ blocks of $\log \beta$ bits are split while the necessary carries are propagated. In order to maintain values within $\mathbb Z_q$, a simple reduction is performed. Figure~\ref{fig:decompose_module} depicts the construction. The resulting $\ell$ polynomials are transformed into the Fourier domain. For this, we instantiate $\ell$ NTT modules, which perform each transform in parallel. The resulting vectors are stored in a ping pong buffer, referred to as \emph{Multi Polynomial Buffer}, allowing computations to be performed while the subsequent polynomials are processed.

\begin{figure*}[htp]
    \centering
    \input{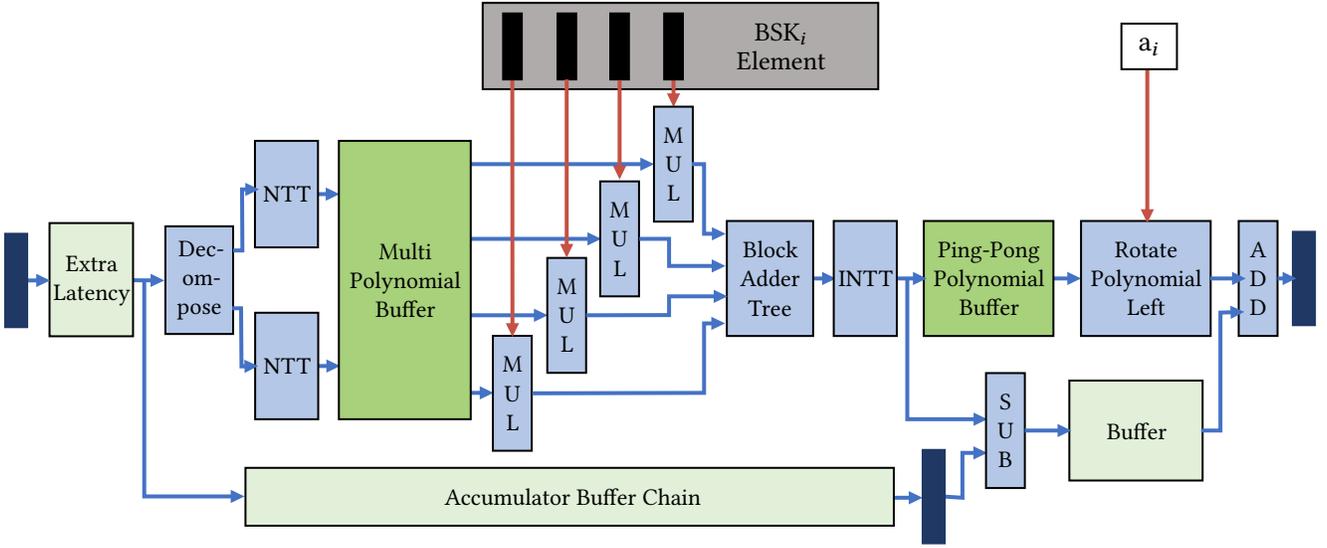}
    \caption{The implementation of a single blind rotation iteration. MUL, Add and Sub denote element-wise multiplication, addition and subtraction of $\mathcal{T}$-many values followed by a modulo reduction.}  \label{fig:blind_rotation_iteration_module}
\end{figure*}

\smallsubsec{Multiplication} In order to multiply two transformed polynomials, we instantiate element-wise multipliers. Each such multiplier performs $\throu$ scalar multiplications in parallel. Thus a polynomial multiplication requires $N/\throu$ iterations. Scaling an RLWE ciphertext by a polynomial is accomplished by repeating the multiplication for each entry in the RLWE ciphertext. Note that during this operation the scaling polynomial remains constant, thus the need to buffer it for the duration of the computation. We instantiate $(k + 1) \times \ell$ element-wise multipliers. This allows us to perform all $(k+1)\times \ell$ polynomial-RLWE products $\decomp(C)_{i,j} \cdot \hatt C_{i,j}$ in parallel. To save resources, we assume that $\bsk_i$ is precomputed, such that it is in NTT space and includes the rescaling factor of the iNTT.

\smallsubsec{Accumulation} In order to efficiently sum the resulting scaled RLWE ciphertexts, we implement a block adder tree. At each level all values are split into pairs and added together, thus halving the number of values. In total $\log((k+1)\times \ell)$ steps are required to perform the entire sum, with additional reductions as needed. Finally, the inverse NTT is applied to all polynomials in the resulting RLWE ciphertext, yielding the final result of the external product.

\subsection{Blind Rotation}
This section details the implementation of the Blind Rotation module, whose block diagram is shown in Figure~\ref{fig:blind_rotation_iteration_module}.
In order to compute the blind rotation, we iterate over the bootstrapping key elements $\bsk_i$ and the LWE mask elements $a_i$.  For each such pair, we compute
$
acc = (acc \boxdot \bsk_i) \cdot (X^{a_i} - 1) + acc.
$
For each iteration $i$, we separate this computation into two steps. We first perform the external product $acc' = acc \boxdot \bsk_i$, which was discussed in the previous section. Then finalize the iteration by computing $acc = acc' \cdot X^{a_i} + acc - acc'$. Note that $acc'$ is an RLWE ciphertext. A multiplication by $X^{a_i}$ is thus equivalent to negacyclically rotating the coefficients in each polynomial in $acc'$ by $a_i$. That is, coefficients with degree $\ge N$ after rotation wrap back around with a negated sign. In order to compute this, we construct a polynomial rotation module. It takes a value $a_i$, as well as a polynomial with coefficients $g_0, \dots, g_{N-1}$. The module rearranges these to $-g_{N-a_i}, \dots, -g_{N-1}, g_0, \dots, g_{N-a_i - 1}$. The negation of a coefficient $g_i$ is computed by $q - g_i$. The negacyclic rotation is performed on all polynomials in the RLWE ciphertext $acc'$. In order to do this, $acc'$ is buffered in a ping pong buffer after the external product. This allows the external product to keep producing subsequent results, while also allowing $acc'$ to be available for the necessary rotations.

Note that it is also possible to perform the polynomial multiplication of $acc' \cdot (X^{a_i} - 1)$ in the NTT domain, potentially accelerating the operation. The values $a_i$ are not available until run time and therefore a real-time transform of $X^{a_i} - 1$ is required. This can be heavily amortized but requires large lookup tables. We instead opt to perform the rotation in the time domain.

During the rotation of $acc'$, we concurrently compute the value $acc - acc'$. Note that for this $acc$ needs to be buffered for the duration of the external product in order to be available, leading to rather large buffers. Finally, we sum the rotated $acc' \cdot X^{a_i}$ with $acc - acc'$ to yield the result of a single blind rotation iteration.
A full blind rotation requires $n$ iterations over the values $a_i \in \vec a$, each using the resulting $acc$ of the previous iteration. Thus, we directly feed the output of the blind rotation back into itself. The final result of a blind rotation on inputs $\bsk$, $acc$, and LWE $(\vec a, b)$ encrypting $\Delta m$ with error $e$ is an RLWE encrypted polynomial $acc \cdot X^{\Delta m + e}$.


\subsection{Programmable Bootstrapping}
The programmable bootstrapping module takes an LWE ciphertext $(\vec a, b)$, an accumulator $acc$ encoding a lookup table, computes the blind rotation given the bootstrapping key, and performs a final sample extraction on the resulting ciphertext.

\begin{figure*}[htp]
    \centering
    \input{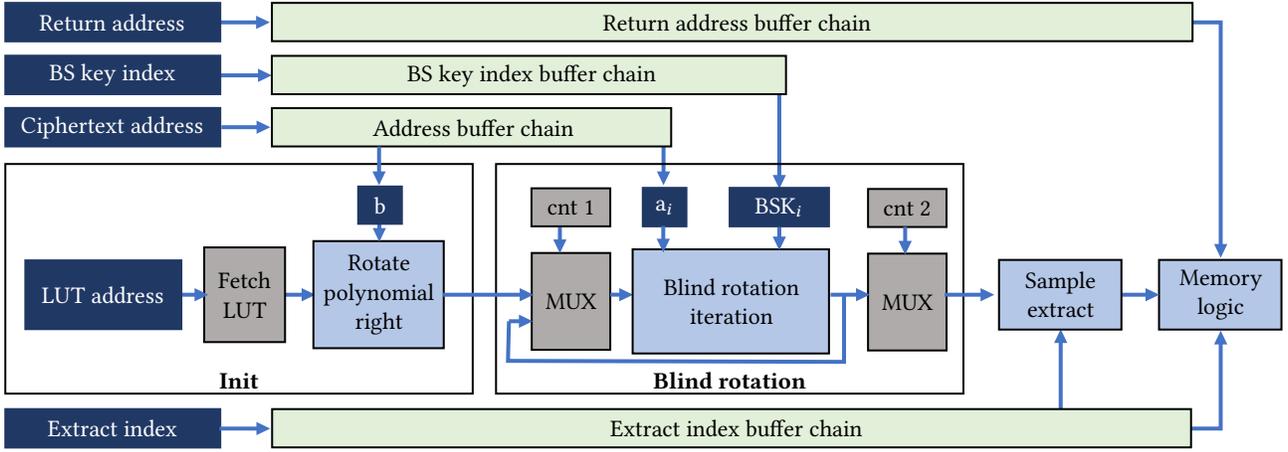}
    \caption{
    PBS module with initialization, blind rotation module and sample extraction. Addresses to the required resources for pipelined operations are provided in buffer chains.}
    \label{fig:pbs_module}
\end{figure*}

Our implementation is able to process multiple ciphertexts at once. For this purpose, we provide the PBS pipeline with an input ciphertext address, the address of the bootstrapping key, the sample extract index, and the final output address. Note that the address of the bootstrapping key is not fixed in cases where different keys are used. While the sample extract index is usually fixed to $0$, our design allows it to be variable. Figure~\ref{fig:pbs_module} illustrates the overview block diagram of our programmable bootstrapping module.

The initialization module takes the body $b$ of the LWE ciphertext to be bootstrapped, as well as the address from which the desired lookup table can be retrieved. Note that the lookup table has the form of an RLWE ciphertext $acc \in \rlwe(F)$ where $F$ is a redundant polynomial. Given $acc$ and $b$ the module negacyclicly rotates all polynomials in $acc$ by $X^{-b}$. We adapt the polynomial rotation module from the blind rotation module for this purpose. 
The sample extract module takes an RLWE ciphertext $C$ and an extraction index $h$. It constructs from this an LWE ciphertext encrypting the coefficient encrypted in $C$ at the given index. In order to accomplish this, the coefficients from $C$ are rearranged as described in Section~\ref{sec:pbs}. This rearrangement is quite similar to a negacyclic polynomial rotation, and therefore, we reuse the polynomial rotation module previously used for the blind rotation for this purpose as well.

\subsection{Key Switching and MulAdd}
\begin{figure}[htp]
    \centering
    \input{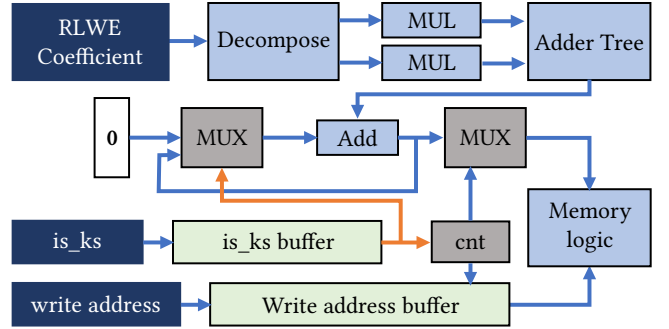}
    \caption{Design of the key-switch and MulAdd module. The \texttt{is\_ks} flag signals whether a key switch or a MulAdd operation is to be performed.}
    \label{fig:ks_module}
\end{figure}

We design a dual-purpose key-switching module that additionally provides fused multiply-add (MulAdd) capabilities. The overview architecture of the dual-purpose key-switching module is illustrated in Figure~\ref{fig:ks_module}.

\smallsubsec{Key-switch module}
The key-switch module takes an LWE ciphertext $(\vec a,b)$ encrypted under a key $\sk$ and a key-switching key $\ksk$ encrypting $\sk$ under $\sk'$. This is accomplished by computing
\[
(\vec 0, b) - \sum_{i=0}^{kN - 1}\sum_{j=1}^\ell\decomp(a_i)_j \cdot \ksk_{i,j}.
\]

Here $\ksk_i$ is a Lev ciphertext consisting of $\ell$ LWE ciphertexts. The product $\decomp(a_i)_j \cdot \ksk_{i,j}$ thus consists of the scaling of the LWE ciphertext $\ksk_{i,j}$ by the $j$-th element of the decomposition of $a_i$. This results in $k + 1$ scalar multiplications per LWE ciphertext.
We instantiate an additional decomposition module as described in Section~\ref{sec:extprod}. Incoming coefficients are decomposed in a pipeline and multiplied with the corresponding $\ksk$ entries. The resulting values are summed in an addition tree and finally returned. We optimize the accumulation of the resulting products by requiring $\ksk_{i,j}$ to be negated. This allows us to perform the sum and subtraction from $(\vec 0, b)$ as a large sum without switching signs.

The key-switching operation is used after the bootstrapping module. In order to ease the pipelining from one module to the next, we aim to match the throughput of one with the other. Note that the key-switching operation is faster than the bootstrapping operation. For a given PBS throughput $\throu_{BS}$ we determine the key-switching throughput to be $\throu_{KS} \geq \frac{\throu_{BS}}{n}$. We pad the input as necessary to ensure $\throu_{KS}$ is not fractional.

\smallsubsec{MulAdd}
The key-switching module is able to produce results faster than the PBS module. We therefore utilize the additional clock cycles of this module to perform additions and multiplications as required. We accomplish this by providing additional flags to deactivate the accumulation and instead directly return results. This allows us to perform MulAdd operations. Subtractions are accomplished by providing the modular representation of a negative scaling factor.

\section{Towards a TFHE processor}

The objective of this work is to advance toward a fully functional and scalable TFHE accelerator architecture, initially prototyped and validated on FPGAs as a foundation for future ASIC implementations. This approach bypasses the severe bandwidth constraints incurred when only selected operations are offloaded to the FPGA.

Our current implementation supports the programmable bootstrapping operation. Additional infrastructure is implemented in order to make the module programmable. That is, the module is provided with data addresses to retrieve the appropriate objects from memory. This contrasts with prior implementations wherein the module receives all data from outside. This design step allows us to implement a custom instruction set, which can be interpreted and executed by the FPGA, instead of the FPGA having a single accelerated functionality.

Currently, we do not provide an implementation of a key-switch module. As our design assigns the MulAdd operation to this module, these operations are currently not available as well. Our design however, demonstrates the feasibility of an implementation of the dual-purpose key-switching module alongside the already implemented PBS module. A subsequent step requires the implementation of a custom instruction, as well as logic for its processing. 

Note that our design does not include a general external product for ciphertext-ciphertext multiplication. Instead, we assume that homomorphic multiplications are performed by concatenating messages and performing a programmable bootstrap with an accordingly constructed lookup table. This is also the approach followed by the tfhe-rs library~\cite{tfhe_rs}.

\subsection{External Bandwidth Requirements}
With all the optimizations discussed in Section~\ref{sec:implementation}, our design makes extensive use of the FPGA’s on-chip memory. However, certain operators and data objects required during circuit evaluation exceed on-chip capacity and must be stored off-chip. Consequently, the design imposes external bandwidth requirements that must be met to ensure correct operation at the target frequency.

In general, the external bandwidth of the TFHE processor must at least support transferring PBS/s-many operations per second with an additional margin to transfer the bootstrapping key, key-switch key, and lookup table, as well as the necessary ciphertexts. 

Assuming a 64-bit address space, a processor instruction operation requires $3 \cdot 64$ bits for the data addresses and $\log N$ bits specifically required for sample extraction. Finally, additional bits are required for the instructions themselves. Two bits suffice for this, assuming the minimal instruction set PBS, AddMul, and KS. Note that the keys are stored on the FPGA, saving additional external bandwidth requirements.
Considering the largest parameter sets tested, as well as the highest rate of PBSs per second achieved by our design, this results in an external bandwidth requirement of around 3 Megabits per second. Prior work requires several Gigabytes per second~\cite{ALT,Tian_Ye,Tianqi_Kong}.

\subsection{Internal Bandwidth Requirements}
The highest internal bandwidth requirement incurred by our design is from the bootstrapping key BSK. During bootstrapping, different parts of the key are required. The underlying RGSW ciphertexts are quite considerable in size.
Loading of the bootstrapping key is amortized through batched bootstrapping as proposed in FPT~\cite{FPT}. This keeps a key element longer in use, allowing more clock cycles for the subsequent element to be loaded from memory. Setting the batch size trades lower memory consumption and latency for a higher internal memory bandwidth requirement.

Assuming the highest performing configuration accomplished by our design ($\throu = 32$ with frequency $f = 325$ MHz) and the benchmark parameter set $k = 1, \ell = 2, n = 500$, this yields a minimal batchsize of $9$. This results in an internal bandwidth requirement of around $93$ GB/s for the PBS module, as well as an additional $73$ GB/s for the key-switch module. This totals $166$ GB/s, which has to be met by the underlying hardware. 

\section{Experiments and Results}\label{sec:results}

In this section, we discuss the results of our proposed hardware-accelerated TFHE processor. Specifically, first we introduce the experimental setup, then we introduce the baseline configurations that are used in the evaluations to facilitate fair comparison, next we present the results in terms of hardware overhead, NTT performance, and full system performance.

\subsection{Experimental Setup}\label{subsubsec:system}

\begin{figure}[htp]
    \centering
    \input{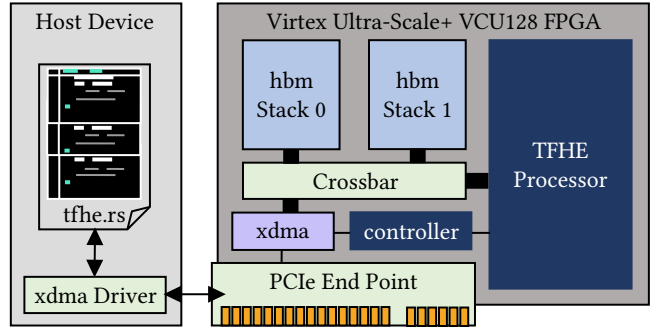}
    \caption{System-level integration of the proposed TFHE processor implemented on the AMD Virtex UltraScale+ HBM VCU128 FPGA. The host device and FPGA communicate via memory-mapped I/O (MMIO) over a PCIe interface.}
    \label{fig:full_system}
\end{figure}

\begin{table*}[htp]
\caption{Resources of the FPGAs used by different implementations. All FPGAs additionally feature multiple GB off-chip DDR4 RAM. SRAM is BRAM + URAM + PLRAM. RAM unit conversions: 1 LUTRAM = 64 bit, 1 BRAM = 36 kb, 1 URAM = 288 kb. *LUTRAM for the FPGA used by YKP is estimated based on the fact that one half of CLB LUTs are LUTRAM~\cite{vivado_documentation_lutram} and the VU13P has 1728 K CLB LUTs~\cite{vu13_datasheet}. Abbreviations: DCAC (Data Center Accelerator Card), VU+ (Virtex UltraScale+ HBM)}
    \label{tab:fpga_comparison}
    \centering
    \begin{tabular}{c|c|c|c|c|c|c|c|c|c|c}
          & FPGA & LUT & FF & DSP & LUTRAM & BRAM & URAM & PLRAM & SRAM & Other \\ \hline
          \\[-1em]
          FPT~\cite{FPT} & \pbox{1.6cm}{AMD Alveo U280 DCAC} & 1304 K & 2607 K & 9024 & N/A & 2016 & 960 & 24 Mb & 365 Mb & \pbox{3cm}{8 GB HBM with up to 460 GB/s}\\[8pt] \hline
          \\[-1em]
          ALT~\cite{ALT} & \pbox{1.6cm}{AMD Alveo U250 DCAC} & 1728 K & 3456 K & 12288 & N/A & 2688 & 1280 & N/A & 452 Mb & \pbox{3cm}{External memory with up to 77 GB/s}\\[8pt] \hline
          \\[-1em]
          KL~\cite{Tianqi_Kong} & \pbox{1.6cm}{AMD VU+ VCU128} & 1304 K & 2607 K & 9024 & 601 K & 2016 & 960 & 0 Mb & 341 Mb & \pbox{3cm}{8 GB HBM with up to 460 GB/s}\\[8pt] \hline
          \\[-1em]
          YKP~\cite{Tian_Ye} & \pbox{1.6cm}{AMD VU+ VU13P} & 1728 K & 3456 K & 12288 & 864 K* & 2688 & 1280 & 0 Mb & 452 Mb & \pbox{3cm}{External memory with up to 77 GB/s}\\[8pt] \hline
          \\[-1em]
          Ohba~\cite{Ohba2025_nvme} &  \pbox{1.6cm}{AMD XCVU47P} & 1304 K & 2607 K & 9024 & 601 K & 2016 & 960 & 0 Mb & 341 Mb & \pbox{3cm}{16 GB HBM with up to 460 GB/s}\\[8pt] \hline
          \\[-1em]
          This paper & \pbox{1.6cm}{AMD VU+ VCU128} & 1304 K & 2607 K & 9024 & 601 K & 2016 & 960 & 0 Mb & 341 Mb & \pbox{3cm}{8 GB HBM with up to 460 GB/s}\\[5pt]
    \end{tabular}
    
\end{table*}

Our design requires a high-end FPGA equipped with High Bandwidth Memory (HBM) and an integrated PCIe interface. We evaluate our implementation on an AMD Virtex UltraScale+ VCU128 FPGA, with two HBM stacks, and a high-speed PCI Express (PCIe) interface with 16 GT/s. The overall system architecture of our proposed THFE processor is illustrated in Figure~\ref{fig:full_system}. 

The FPGA communicates with the host device through a PCIe Gen2 x8 interface using the XDMA IP core, which provides bidirectional data transfer between host memory and on-chip memory via memory-mapped I/O (MMIO). On the host side, the modified THFE library (tfhe.rs) communicates to the xdma driver, which facilitates low-latency communication to the FPGA hardware.
Within the FPGA, the PCIe Endpoint interfaces with the XDMA core, which manages DMA channels for efficient host-device data transfers. A custom controller module coordinates command and data transactions between the XDMA core and the on-chip crossbar interconnect. The crossbar enables concurrent access to both HBM stacks, allowing the TFHE processor to achieve high throughput for data-intensive homomorphic operations.

The TFHE processor itself is tightly coupled to the crossbar, providing direct access to the high-bandwidth HBM memory (up to 460 GB/s aggregate bandwidth on the VCU128). Each HBM stack is organized into multiple pseudo-channels to support parallel data streams, reducing latency and improving concurrency. This configuration enables efficient evaluation of gate-level homomorphic operations, where large ciphertexts must be frequently read and written during bootstrapping and gate execution.

\subsection{Baseline for Comparisons}\label{subsec:res-baseline}

To compare the hardware resource utilization of our implementation with existing work, we first establish a baseline configuration, as direct comparison is non-trivial due to variations in hardware platforms, implemented operations, and parameter choices. Therefore, we define a baseline over the blind rotation module, as it is the heaviest operation and is provided by most prior works. Specifically, we scale prior 32-bit implementations by a factor of two to account for the increased computational cost of 64-bit operations used in our design. This adjustment favors 32-bit architectures, as the cost of arithmetic operations grows exponentially with operand precision.

While prior works employ slightly different Boolean parameter sets, typically $n = 500$, $N = 1024$, $k = 1$, $\ell = 2$, $\log \beta = 10$, and $q = 2^{32}$, ALT adopts modified parameters with $\ell = 1$, $\log \beta = 16$, and a composite modulus. To enable fair comparison, we normalize the decryption failure rate (DFR) by assuming that repeating a computation squares the DFR. Accordingly, we scale it by
$z = \frac{\log_2(\text{DFR}_x)}{-64}$
to match the standard DFR of $2^{-64}$~\cite{concrete_library}.
This normalization inherently compensates for parameter variations that affect the DFR.

\begin{table*}[htp]
 \caption{Performance comparison of different 1024-point NTT designs. All values are rounded. A dash denotes missing data. Latency and clock cycles are computed using the relationship $latency\cdot frequency=clock\text{ } cycles$ where either is not provided by the authors. The results for our NTT were measured with a negacyclic NTT using $\mathcal{T}=2$.}\label{tab:ntt_comparison}
    \centering
    \begin{tabular}{c|c|c|c|c|c|c|c|c|c|c}
         Design & $\lceil log_2(p) \rceil$ & \pbox{1cm}{Clock\newline /MHz} & \pbox{1.4cm}{Clock\newline cycles} & \pbox{1.25cm}{Delay\newline /µs} & NTTs/ms & LUT & FF & DSP & LUTRAM & BRAM\\ \hline
         HLW~\cite{MQT} & 64 & 179 & 1294 & 7.23 & 138 & 9 K & 9 K & 128 & - & 8 \\
         KL~\cite{Tianqi_Kong} & 52 & 433 & 697 & 1.61 & 621 & 33 K & 44 K & 160 & - & 40 \\
         YYKKP~\cite{Tian_Ye_NTT} & 32 & 215 & \textbf{198} & \textbf{0.92} & 1086 & 94 K & 105 K & 640 & - & 80 \\
         Proteus~\cite{Proteus} & 32 & 150 & 1100 & 7.3 & 300 & 7 K & 3 K & 36 & - & 2 \\
         CM~\cite{Ahmed_Can_Mert_2020} & 28 & 125 & 490 & 3.90 & 255 & 16 K & 14 K & 56 & - & 24 \\
         This paper & 64 & \textbf{600} & 700 & 1.17 & \textbf{1172} & 12 K & 19 K & 90 & 980 & 8 \\
    \end{tabular}
\end{table*}

\subsection{Hardware Overhead}\label{subsec:res-overhead}

In this section, we analyze and compare the resource utilization of the proposed TFHE processor against state-of-the-art FPGA-based implementations. Although prior works target different FPGA platforms, we consider them broadly comparable since all devices originate from the same manufacturer and exhibit similar resource architectures and performance characteristics. All implementations operate well below their respective resource limits. Table~\ref{tab:fpga_comparison} summarizes the comparison in terms of FPGA resources.

\subsection{NTT Performance Comparison}\label{subsec:res-ntt-performance}
In order to illustrate the efficiency of our implementation, we compare our 64-bit negacyclic NTT with prior FPGA-based designs, summarized in Table~\ref{tab:ntt_comparison}. Among them, YYKKP~\cite{Tian_Ye} and KL~\cite{Tianqi_Kong} denote the NTT implementations of the respective TFHE accelerators, while Proteus~\cite{Proteus}, HLW~\cite{MQT}, and CM~\cite{Ahmed_Can_Mert_2020} provide standalone 1024-point NTTs. All use Virtex-family FPGAs with comparable resource architectures.
Our design, operating at 600 MHz with $\mathcal{T}=2$, achieves the highest frequency, throughput (NTTs/ms) and lowest latency among all works except for YYKKP, which attains slightly lower latency but requires over $7\times$ more DSPs, $5\times$ more FFs, and $7\times$ more LUTs while using half the bit width. Compared to HLW, which uses the same modulus $p=2^{64}-2^{32}+1$, our design delivers $\sim9\times$ higher throughput and $6\times$ lower latency. It also outperforms KL on the same FPGA and exceeds Proteus and CM by around $4\times$ in performance while maintaining competitive resource efficiency.
Overall, our NTT demonstrates superior performance with resource balance and forms a high-speed backbone for polynomial multiplications within the blind rotation.

\subsection{Full System Performance Comparison}\label{subsec:res-performance}

In this section, we compare the performance of the proposed TFHE processor against state-of-the-art implementations using the baseline configurations described in Section~\ref{subsec:res-baseline}. In order to have a fair comparison between prior works that utilize standard and large parameter sets, we conducted separate experiments.

\smallsubsec{Evaluation on Standard Parameter Set} Table~\ref{tab:pbs_performance_comparison} summarizes the results in terms of decryption failure rate (DFR), number of coefficients processed per clock cycle ($\mathcal{T}$), operating frequency, throughput in PBS/s (for 64-bit, 32-bit, and 32-bit normalized to DFR $2^{-64}$), latency, hardware resource utilization, and power consumption. For the sake of completeness, CPU and GPU implementations are also included as reference baselines in the comparison. Note that we take the DFR for ALT and FPT as reported by ALT~\cite{ALT}.

\begin{table*}[ht]
\caption{
    Comparison with literature results on the parameter set: $n=500$, $N=1024$, $k=1$, $\ell=2$ and $\log \beta = 10$. * denotes estimated data. The comparison CPU is an AMD 7950X3D running tfhe-rs~\cite{tfhe_rs}.
    The comparison GPU 1 is an Nvidia L40 running tfhe-rs using the classic settings. 
    The comparison GPU 2 is an RTX 3090 running cuFHE~\cite{cuFHE}, its results are from YKP's benchmark~\cite{Tian_Ye}.
    For the CPU and GPU 1 data, we took the Thermal Design Power (TDP) as their power consumption and averaged their performance over $10^4$ runs.
    }
    \label{tab:pbs_performance_comparison}
    \centering
    \begin{tabular}{c|c|c|c|c|c|c|c|r r r r c}
          & DFR & $\mathcal{T}$ & \pbox{1cm}{Clock\newline /MHz} & \pbox{1cm}{PBS/s \newline 64-bit} &  \pbox{1cm}{PBS/s \newline 32-bit} & \pbox{2cm}{PBS/s 32-bit \newline with DFR $2^{-64}$} & \pbox{1cm}{Delay\newline /ms} & LUT & FF & DSP & SRAM & \pbox{1.6cm}{Power/W, PBS per W}\\[5pt] \hline
          \\[-1em]
         FPT~\cite{FPT} & $2^{-15}$ & \textbf{128} & 200 & - & \textbf{25000} & 5860 & 0.74 & 595 K & 1024 K & 5980 & 15 Mb & 99(59)\\[5pt] \hline
         \\[-1em]         
         ALT~\cite{ALT} & $2^{-32}$ & - & 115 & - & 6506 & 3253 & \textbf{0.42} & 623 K & 256 K & 2144 & 74 Mb & 23(141) \\
             &           & - & 152 & - & 6447 & 3224 & 0.46 & 345 K & 162 K & 1120 & 73 Mb & 15(\textbf{215}) \\[5pt] \hline
          \\[-1em]         
         KL~\cite{Tianqi_Kong} & $2^{-64}$ & - & 200 & - & - & - & 2.13 & 414 K & 625 K & 1281 & 40 Mb & - \\
            &           & - & 200 & - & *699 & *699 & 1.43 & 510 K & 750 K & 1281 & 56 Mb & - \\[5pt] \hline
          \\[-1em]         
         YKP~\cite{Tian_Ye} & $2^{-64}$ & - & 180 & - & 3454 & 3454 & 3.76 & 842 K & 662 K & 7202 & 338 Mb & 50(69) \\
             &           & - & 180 & - & 2657 & 2657 & 1.88 & 442 K & 342 K & 6910 & 409 Mb & 50(53) \\[5pt] \hline
          \\[-1em]         
         This paper & $2^{-64}$ & 2  & \textbf{575} & 1123 & 2246 & 2246 & 2.67 &  61 K &   92 K &  342 &  3 Mb & 16(140) \\
                    &           & 4  & 525 & 2050 &  4100 &  4100 & 1.95 &  112 K &  177 K &  684 &  4 Mb & 27(152) \\
                    &           & 8  & 450 & 3516 &  7032 &  7032 & 1.42 & 201 K &  352 K & 1368 &  6 Mb & 42(167) \\                    
                    &           & 16 & 400 & 6250 & 12500 & 12500 & 0.96 & 388 K &  697 K & 2736 &  10 Mb & 73(171) \\                    
                    &           & 32 & 325 & 10156 & 20312 & \textbf{20312} & 0.88 & 781 K & 1391 K & 5472 & 14 Mb & 119(171) \\ \hline \hline
          \\[-1em]
         CPU 16 core & $2^{-64}$ & -  & 4200 & 3513 & 7026 &  7026 & 4.21 &  - & - &  - &  - & 120(59) \\
        CPU 1 core & $2^{-64}$ & -  & 5700 & 237  & 474  &  474  & 4.21  &  - & - &  - &  - & 120 \\
         GPU 1 & $2^{-64}$ & -  & 2500 & 8689 & 17378 & 17378 & 23.03 &  - & - &  - &  - & 300(58) \\ 
         GPU 2\cite{Tian_Ye} & $2^{-64}$ & -  & 1700 & -  & 9600 & 9600 & 9.34 &  - & - &  - &  - & 350(27) \\
    \end{tabular}
    
\end{table*}

When accounting for DFR, our implementation with $\mathcal{T}=32$ achieves over $3.4\times$ higher throughput than the next fastest design (FPT) and more than $5.8\times$ over other FPGA implementations, while maintaining comparable total resource use. For a fair performance-per-resource comparison, consider the $\mathcal{T}=4$ measurement, which achieve similar PBS/s to ALT, KL, and YKP while requiring far fewer resources and delivering 18\% higher throughput. 

In terms of energy efficiency, our processor achieves the second-best performance per watt, trailing only ALT. Regarding latency, our best configuration ranks behind ALT and FPT. Note that our 64-bit implementation is at a disadvantage since 32-bit multiplications require fewer cycles than our 64-bit multiplications. However, ALT attains the lowest latency primarily because of its smaller multiplications due to their composite NTT and additionally through employing bootstrapping key unrolling, which also reduces computation effort~\cite{ALT}. Incorporating key unrolling in future versions could further improve our design.
Our best performance-per-resource ratio is achieved at $\mathcal{T}=2$, while the highest absolute throughput occurs at $\mathcal{T}=32$. As $\mathcal{T}$ increases, frequency decreases due to longer interconnecting paths and higher routing congestion. Notably, a standalone NTT with $\mathcal{T}=8$ reaches nearly 600 MHz, indicating that inter-block routing rather than the NTT itself limits maximum frequency.

\begin{table*}[ht]
  \caption{
    Comparison with literature results on the parameter set: $n=800$, $N=16384$, $k=1$, $\ell=5$ and $\log \beta =6$ and DFR $2^{-64}$. * denotes estimated data.
    The comparison CPU is an AMD 7950X3D running TFHE-rs~\cite{tfhe_rs} using all threads.
    The comparison GPU is an Nvidia L40 running TFHE-rs using the classic settings.
    }
    \label{tab:pbs_performance_comparison_big_parameterset}
    \centering
    \begin{tabular}{c|c|c|c|c|r r r r r r c}
          & $\mathcal{T}$ & \pbox{1cm}{Clock\newline /MHz} & \pbox{1cm}{PBS/s \newline 64-bit} & \pbox{1cm}{Delay\newline /ms} & LUT & FF & DSP & LUTRAMs & BRAMs & URAMs & \pbox{1.6cm}{Power/W, PBS per W}\\[5pt] \hline
          \\[-1em]
         Ohba~\cite{Ohba2025_nvme} & - & 200 & *4 & 250 & 461 K & 615 K & 1549 & - & 1024 & 64 & -\\[5pt] \hline       
         This paper & 8  & \textbf{400} & 122 & 25 &  529 K &   939 K &  3744 & 55 K &  662 & 128 & 105(1.16) \\
         \hline \hline
          \\[-1em]
         CPU 16 core & -  & 4200 & 28 &  314 &  - & - & - &  - &  - & - & 120(0.23) \\
        GPU & -  & 2500 & 109 & 1834 &  - & - & - &  - &  - & - & 300 \\ 
    \end{tabular}
  
\end{table*}

\smallsubsec{Evaluation on Large Parameter Set} To demonstrate scalability and flexibility, we evaluate our TFHE processor using the demanding parameter set from Ohba et al.~\cite{Ohba2025_nvme}, as shown in Table~\ref{tab:pbs_performance_comparison_big_parameterset}. Similar to previous experiments, we have included the CPU and GPU results for reference. Our design achieves over $30\times$ higher throughput and $10\times$ lower latency. Compared to a 16-core CPU (GPU) baseline, it delivers a $2.89\times$ ($1.16\times$) speedup on the smaller parameter set and $4.35\times$ ($1.12\times$) on the larger one, showing improved efficiency at scale against the CPU, and similar scalability to the L40 GPU.

In our architecture, parameter $k$ affects only buffer sizes, while $N$ primarily impacts BRAM and URAM blocks already partitioned into $\mathcal{T}$ sub-buffers. As long as these fit within FPGA RAM primitives, scaling parameters proportionally scale performance, such that they have a negligible performance impact.
Technological differences also affect cross-platform comparisons: our FPGA is manufactured using a 20~nm process~\cite{virtex_ultrascale_fpga_product_page}, while comparison CPU and GPUs rely on 5–6 nm FinFET nodes~\cite{7950x3d_datasheet,L40_datasheet}. Despite this disadvantage, our $\mathcal{T}=32$ design achieves over $5\times$ ($28\times$) lower latency and $2.8\times$ ($1.16\times$) higher throughput than a 16-core CPU (GPU) with similar or lower power consumption.
In summary, these results highlight that our FPGA accelerator surpasses general-purpose processors in performance and efficiency, highlighting strong potential for ASIC-based TFHE solutions and practical privacy-preserving computation in real-world applications.

\section{Conclusion and future work}

TFHE is a fast, torus-based fully homomorphic encryption scheme that supports both linear and non-linear operations, offering the most efficient bootstrapping among FHE schemes. However, homomorphic circuit evaluation still incurs significant computational overhead compared to unencrypted computation. As a solution, prior efforts have accelerated individual TFHE operations, particularly the blind rotation, using FPGAs. Yet, these approaches suffer from severe bandwidth limitations caused by frequent data transfers between the host and the FPGA.

Our work addresses this limitation by proposing a fully integrated TFHE accelerator implemented entirely on the FPGA. The design incorporates all essential modules for TFHE functionality and supporting infrastructure to execute complete homomorphic circuits directly on the hardware, effectively eliminating off-chip bandwidth bottlenecks.
The current prototype includes designs for addition, scalar multiplication, programmable bootstrapping, and key switching, with the bootstrapping module already implemented. Results demonstrate a compact and flexible design that significantly outperforms existing approaches while supporting diverse parameter sets.

Future work includes completing the key-switching and instruction modules, defining a dedicated instruction set, and developing a compiler to translate homomorphic circuits into executable hardware instructions. Additional optimizations, such as parameter-specific tuning, key unrolling, and composite NTT implementations, could further enhance performance.

In conclusion, our proposed TFHE accelerator demonstrates strong potential for scalable FPGA implementations and establishes a foundation for future dedicated FHE processor architectures and ASIC hardware.

\bibliographystyle{ACM-Reference-Format}
\bibliography{bibfile}

@inproceedings{Regev2005,
  series = {STOC05},
  title = {On lattices,  learning with errors,  random linear codes,  and cryptography},
  url = {http://dx.doi.org/10.1145/1060590.1060603},
  DOI = {10.1145/1060590.1060603},
  booktitle = {Proceedings of the thirty-seventh annual ACM symposium on Theory of computing},
  publisher = {ACM},
  author = {Regev,  Oded},
  year = {2005},
  month = may,
  collection = {STOC05}
}

@article{lyubashevsky_ideal_2012,
  author       = {Vadim Lyubashevsky and
                  Chris Peikert and
                  Oded Regev},
  title        = {On Ideal Lattices and Learning with Errors over Rings},
  journal      = {J. {ACM}},
  volume       = {60},
  number       = {6},
  pages        = {43:1--43:35},
  year         = {2013},
  url          = {https://doi.org/10.1145/2535925},
  doi          = {10.1145/2535925},
  bibsource    = {dblp computer science bibliography, https://dblp.org}
}

@Misc{tfhe_rs,
  title={{TFHE-rs: A Pure Rust Implementation of the TFHE Scheme for Boolean and Integer Arithmetics Over Encrypted Data}},
  author={Zama},
  year={2022},
  note={\url{https://github.com/zama-ai/tfhe-rs}},
}

@article{Tianqi_Kong,
  title = {Hardware Acceleration and Implementation of Fully Homomorphic Encryption Over the Torus},
  volume = {71},
  ISSN = {1558-0806},
  url = {http://dx.doi.org/10.1109/TCSI.2023.3338953},
  DOI = {10.1109/tcsi.2023.3338953},
  number = {3},
  journal = {IEEE Transactions on Circuits and Systems I: Regular Papers},
  publisher = {Institute of Electrical and Electronics Engineers (IEEE)},
  author = {Kong,  Tianqi and Li,  Shuguo},
  year = {2024},
  month = mar,
  pages = {1116–1129}
}

@inproceedings{Tian_Ye,
  title = {FPGA Acceleration of Fully Homomorphic Encryption over the Torus},
  url = {http://dx.doi.org/10.1109/HPEC55821.2022.9926381},
  DOI = {10.1109/hpec55821.2022.9926381},
  booktitle = {2022 IEEE High Performance Extreme Computing Conference (HPEC)},
  publisher = {IEEE},
  author = {Ye,  Tian and Kannan,  Rajgopal and Prasanna,  Viktor K.},
  year = {2022},
  month = sep,
  pages = {1–7}
}

@inproceedings{FPT,
  series = {CCS ’23},
  title = {FPT: A Fixed-Point Accelerator for Torus Fully Homomorphic Encryption},
  url = {http://dx.doi.org/10.1145/3576915.3623159},
  DOI = {10.1145/3576915.3623159},
  booktitle = {Proceedings of the 2023 ACM SIGSAC Conference on Computer and Communications Security},
  publisher = {ACM},
  author = {Van Beirendonck,  Michiel and D’Anvers,  Jan-Pieter and Turan,  Furkan and Verbauwhede,  Ingrid},
  year = {2023},
  month = nov,
  pages = {741–755},
  collection = {CCS ’23}
}

@article{ALT,
  title = {ALT: Area-Efficient and Low-Latency FPGA Design for Torus Fully Homomorphic Encryption},
  volume = {32},
  ISSN = {1557-9999},
  url = {http://dx.doi.org/10.1109/TVLSI.2024.3353374},
  DOI = {10.1109/tvlsi.2024.3353374},
  number = {4},
  journal = {IEEE Transactions on Very Large Scale Integration (VLSI) Systems},
  publisher = {Institute of Electrical and Electronics Engineers (IEEE)},
  author = {Hu,  Xiao and Li,  Zhihao and Wang,  Zhongfeng and Lu,  Xianhui},
  year = {2024},
  month = apr,
  pages = {645–657}
}

@inproceedings{gentry09,
author = {Gentry, Craig},
title = {Fully homomorphic encryption using ideal lattices},
year = {2009},
isbn = {9781605585062},
publisher = {Association for Computing Machinery},
address = {New York, NY, USA},
url = {https://doi.org/10.1145/1536414.1536440},
doi = {10.1145/1536414.1536440},
booktitle = {Proceedings of the Forty-First Annual ACM Symposium on Theory of Computing},
pages = {169–178},
numpages = {10},
location = {Bethesda, MD, USA},
series = {STOC '09}
}

@article{MXP,
    title = {An FPGA-based Programmable Vector Engine for Fast Fully Homomorphic Encryption over the Torus},
    url = {https://par.nsf.gov/biblio/10282639},
    abstractNote = {This paper describes an FPGA-based vector engine to accelerate the bootstrapping procedure of Fast Fully Homomorphic Encryption over the Torus (TFHE), a popular and high-performance fully homomorphic encryption scheme. Most TFHE bootstraping comprises many matrix-vector operations that are implemented using Torus polynomials, which are not efficiently implemented on today's standard arithmetic hardware. Our implementation achieves linear performance scaling with up to 16 vector lanes. Future work will switch to an FFT-based polynomial multiplication scheme and switch to larger FPGA parts to accommodate more vector lanes.},
    journal = {SPSL: Secure and Private Systems for Machine Learning (ISCA Workshop)},
    author = {Gener, Serhan and Newton, Parker and Tan, Daniel and Richelson, Silas and Lemieux, Guy and Brisk, Philip},
    year = {2021}
}

@InProceedings{barrett,
author="Barrett, Paul",
editor="Odlyzko, Andrew M.",
title="Implementing the Rivest Shamir and Adleman Public Key Encryption Algorithm on a Standard Digital Signal Processor",
booktitle="Advances in Cryptology --- CRYPTO' 86",
year="1987",
publisher="Springer Berlin Heidelberg",
address="Berlin, Heidelberg",
pages="311--323",
isbn="978-3-540-47721-1"
}

@article{TFHE_original_paper,
  title = {TFHE: Fast Fully Homomorphic Encryption Over the Torus},
  volume = {33},
  ISSN = {1432-1378},
  url = {http://dx.doi.org/10.1007/s00145-019-09319-x},
  DOI = {10.1007/s00145-019-09319-x},
  number = {1},
  journal = {Journal of Cryptology},
  publisher = {Springer Science and Business Media LLC},
  author = {Chillotti,  Ilaria and Gama,  Nicolas and Georgieva,  Mariya and Izabachène,  Malika},
  year = {2019},
  month = apr,
  pages = {34–91}
}

@misc{vivado_documentation_lutram,
      author = {AMD},
      title = {Versal Adaptive SoC Configurable Logic Block Architecture Manual (AM005)},
      year = {2024},
      note = {\url{https://docs.amd.com/r/en-US/am005-versal-clb/LUTRAM}},
      url = {https://docs.amd.com/r/en-US/am005-versal-clb/LUTRAM}
}

@misc{vu13_datasheet,
      author = {AMD},
      title = {AMD UltraScale+ FPGAs Product Selection Guide (XMP103)},
      year = {2024},
      note = {\url{https://docs.amd.com/v/u/en-US/ultrascale-plus-fpga-product-selection-guide}},
      url = {https://docs.amd.com/v/u/en-US/ultrascale-plus-fpga-product-selection-guide}
}

@misc{L40_datasheet,
      author = {techpowerup.com},
      title = {NVIDIA L40},
      year = {2025},
      note = {\url{https://www.techpowerup.com/gpu-specs/l40.c3959}},
      url = {https://www.techpowerup.com/gpu-specs/l40.c3959}
}

@misc{7950x3d_datasheet,
      author = {AMD},
      title = {AMD Ryzen 9 7950X3D Gaming Processor},
      year = {2025},
      note = {\url{https://www.amd.com/en/products/processors/desktops/ryzen/7000-series/amd-ryzen-9-7950x3d.html}},
      url = {https://www.amd.com/en/products/processors/desktops/ryzen/7000-series/amd-ryzen-9-7950x3d.html}
}

@misc{virtex_ultrascale_fpga_product_page,
      author = {AMD},
      title = {AMD Virtex Ultrascale FPGAs},
      year = {2025},
      note = {\url{https://www.amd.com/de/products/adaptive-socs-and-fpgas/fpga/virtex-ultrascale.html}},
      url = {https://www.amd.com/de/products/adaptive-socs-and-fpgas/fpga/virtex-ultrascale.html}
}

@article{Proteus,
  title = {Proteus: A Pipelined NTT Architecture Generator},
  volume = {32},
  ISSN = {1557-9999},
  url = {http://dx.doi.org/10.1109/TVLSI.2024.3377366},
  DOI = {10.1109/tvlsi.2024.3377366},
  number = {7},
  journal = {IEEE Transactions on Very Large Scale Integration (VLSI) Systems},
  publisher = {Institute of Electrical and Electronics Engineers (IEEE)},
  author = {Hirner,  Florian and Mert,  Ahmet Can and Roy,  Sujoy Sinha},
  year = {2024},
  month = jul,
  pages = {1228–1238}
}

@article{ntt_guide,
  title = {Conceptual Review on Number Theoretic Transform and Comprehensive Review on Its Implementations},
  volume = {11},
  ISSN = {2169-3536},
  url = {http://dx.doi.org/10.1109/ACCESS.2023.3294446},
  DOI = {10.1109/access.2023.3294446},
  journal = {IEEE Access},
  publisher = {Institute of Electrical and Electronics Engineers (IEEE)},
  author = {Satriawan,  Ardianto and Syafalni,  Infall and Mareta,  Rella and Anshori,  Isa and Shalannanda,  Wervyan and Barra,  Aleams},
  year = {2023},
  pages = {70288–70316}
}

@inproceedings{Ahmed_Can_Mert_2020,
  author={Mert, Ahmet Can and Karabulut, Emre and Öztürk, Erdinç and Savaş, Erkay and Becchi, Michela and Aysu, Aydin},
  booktitle={{2020 Design, Automation \& Test in Europe Conference \& Exhibition (DATE)}}, 
  title={A Flexible and Scalable NTT Hardware : Applications from Homomorphically Encrypted Deep Learning to Post-Quantum Cryptography}, 
  year={2020},
  volume={},
  number={},
  pages={346-351},
  doi={10.23919/DATE48585.2020.9116470}
}

@inbook{Tian_Ye_NTT,
  title = {FPGA Acceleration of Number Theoretic Transform},
  ISBN = {9783030787134},
  ISSN = {1611-3349},
  url = {http://dx.doi.org/10.1007/978-3-030-78713-4_6},
  DOI = {10.1007/978-3-030-78713-4_6},
  booktitle = {High Performance Computing},
  publisher = {Springer International Publishing},
  author = {Ye,  Tian and Yang,  Yang and Kuppannagari,  Sanmukh R. and Kannan,  Rajgopal and Prasanna,  Viktor K.},
  year = {2021},
  pages = {98–117}
}

@article{Ohba2025_nvme,
  title = {An NVMe-Based Secure Computing Platform With FPGA-Based TFHE Accelerator},
  volume = {13},
  ISSN = {2169-3536},
  url = {http://dx.doi.org/10.1109/ACCESS.2025.3561728},
  DOI = {10.1109/access.2025.3561728},
  journal = {IEEE Access},
  publisher = {Institute of Electrical and Electronics Engineers (IEEE)},
  author = {Ohba,  Yoshihiro and Sanuki,  Tomoya and Gravel,  Claude and Mihara,  Kentaro and Wakasugi,  Asuka and Adachi,  Kenta},
  year = {2025},
  pages = {69980–69997}
}

@inproceedings{MQT,
  title={Number Theoretic Transform (NTT) FPGA Accelerator},
  author={Austin Hartshorn and Humberto Leon and Noel Qiao and Scott J. Weber and Dr. Yarkin Doroz},
  year={2020},
  url={https://api.semanticscholar.org/CorpusID:235498358}
}

@inbook{cuFHE,
  title = {cuHE: A Homomorphic Encryption Accelerator Library},
  ISBN = {9783319291727},
  ISSN = {1611-3349},
  url = {http://dx.doi.org/10.1007/978-3-319-29172-7_11},
  DOI = {10.1007/978-3-319-29172-7_11},
  booktitle = {Cryptography and Information Security in the Balkans},
  publisher = {Springer International Publishing},
  author = {Dai,  Wei and Sunar,  Berk},
  year = {2016},
  pages = {169–186}
}

@inproceedings{F1,
  series = {MICRO ’21},
  title = {F1: A Fast and Programmable Accelerator for Fully Homomorphic Encryption},
  url = {http://dx.doi.org/10.1145/3466752.3480070},
  DOI = {10.1145/3466752.3480070},
  booktitle = {MICRO-54: 54th Annual IEEE/ACM International Symposium on Microarchitecture},
  publisher = {ACM},
  author = {Samardzic,  Nikola and Feldmann,  Axel and Krastev,  Aleksandar and Devadas,  Srinivas and Dreslinski,  Ronald and Peikert,  Christopher and Sanchez,  Daniel},
  year = {2021},
  month = oct,
  pages = {238–252},
  collection = {MICRO ’21}
}

@article{BASALISC,
  title = {BASALISC: Programmable Hardware Accelerator for BGV Fully Homomorphic Encryption},
  ISSN = {2569-2925},
  url = {http://dx.doi.org/10.46586/tches.v2023.i4.32-57},
  DOI = {10.46586/tches.v2023.i4.32-57},
  journal = {IACR Transactions on Cryptographic Hardware and Embedded Systems},
  publisher = {Universitatsbibliothek der Ruhr-Universitat Bochum},
  author = {Geelen,  Robin and Van Beirendonck,  Michiel and L. Pereira,  Hilder V. and Huffman,  Brian and McAuley,  Tynan and Selfridge,  Ben and Wagner,  Daniel and Dimou,  Georgios and Verbauwhede,  Ingrid and Vercauteren,  Frederik and Archer,  David W.},
  year = {2023},
  month = aug,
  pages = {32–57}
}

@article{CoFHEE_extended,
  title = {Silicon-Proven ASIC Design for the Polynomial Operations of Fully Homomorphic Encryption},
  volume = {43},
  ISSN = {1937-4151},
  url = {http://dx.doi.org/10.1109/TCAD.2024.3359526},
  DOI = {10.1109/tcad.2024.3359526},
  number = {6},
  journal = {IEEE Transactions on Computer-Aided Design of Integrated Circuits and Systems},
  publisher = {Institute of Electrical and Electronics Engineers (IEEE)},
  author = {Nabeel,  Mohammed and Gamil,  Homer and Soni,  Deepraj and Ashraf,  Mohammed and Gebremichael,  Mizan Abraha and Chielle,  Eduardo and Karri,  Ramesh and Sanduleanu,  Mihai and Maniatakos,  Michail},
  year = {2024},
  month = jun,
  pages = {1924–1928}
}

@book{fpga_vs_asic_basics,
author = {Kuon, Ian and Rose, Jonathan},
title = {Quantifying and Exploring the Gap Between FPGAs and ASICs},
year = {2009},
isbn = {1441907386},
publisher = {Springer Publishing Company, Incorporated},
edition = {1st}
}

@misc{entry_whitepaper,
    url = {https://go.dualitytech.com/hardware-acceleration-of-FHE},
    author = {Ahmad Al Badawi and David Bruce Cousins and Yuriy Polyakov and Kurt Rohloff},
    publisher = {Duality},
    title = {Hardware Acceleration of Fully Homomorphic Encryption: Making Privacy-Preserving Machine Learning Practical },
    year = {2023}
}

@Misc{concrete_library,
  title={{Concrete: TFHE Compiler that converts python programs into FHE equivalent}},
  author={Zama},
  year={2022},
  note={\url{https://github.com/zama-ai/concrete}},
}

@article{Turbo_FHE,
  title = {Turbo-FHE: Accelerating Fully Homomorphic Encryption With FPGA and HBM Integration},
  volume = {42},
  ISSN = {2168-2364},
  url = {http://dx.doi.org/10.1109/MDAT.2025.3527368},
  DOI = {10.1109/mdat.2025.3527368},
  number = {3},
  journal = {IEEE Design \& Test},
  publisher = {Institute of Electrical and Electronics Engineers (IEEE)},
  author = {Nassar,  Hassan and Bauer,  Lars and Henkel,  J\"{o}rg},
  year = {2025},
  month = jun,
  pages = {86–93}
}

@article{jayasena2024directed,
  title={Directed test generation for hardware validation: A survey},
  author={Jayasena, Aruna and Mishra, Prabhat},
  journal={ACM Computing Surveys},
  volume={56},
  number={5},
  pages={1--36},
  year={2024},
  publisher={ACM New York, NY}
}

\end{document}